\documentclass[fleqn,usenatbib]{aa}  
\usepackage{graphicx}
\usepackage{txfonts}
\usepackage{multicol}
\usepackage{color,ulem}
\usepackage{amsmath, epsfig,natbib}
\usepackage{color,ulem}
\usepackage{float}
\usepackage{newtxtext,newtxmath}
\definecolor{webgreen}{rgb}{0,.5,0}
\definecolor{webbrown}{rgb}{.6,0,0}
\usepackage[colorlinks=true,hyperfootnotes=false,%
   breaklinks=true,%
   plainpages=false, bookmarksnumbered, bookmarksopen=true,
  bookmarksopenlevel=1,%
   urlcolor=webbrown, linkcolor=blue, citecolor=blue]{hyperref}
\newcommand{\pc}{\>{\rm pc}}
\newcommand{\kpc}{\mbox{$\>{\rm kpc}$}} 

\newcommand{\Gyr}{\mbox{$\>{\rm Gyr}$}}

\newcommand\degrees{^\circ}
\newcommand{\avg}[1]{\mbox{$\left<{#1}\right>$}}

\begin{document} 

   \title{Bars and boxy/peanut bulges in thin and thick discs}
   \subtitle{III. Boxy/peanut bulge formation and evolution in presence of thick discs}
\titlerunning{b/p formation in thin and thick disc}
   \author{
   Soumavo Ghosh\inst{1} \thanks{E-mail: ghosh@mpia-hd.mpg.de},
   Francesca Fragkoudi \inst{2},
   Paola Di Matteo \inst{3}
   \and
   Kanak Saha \inst{4}
   }
\authorrunning{S. Ghosh et al.}
   \institute{Max-Planck-Institut f\"{u}r Astronomie, K\"{o}nigstuhl 17, D-69117 Heidelberg, Germany
   \and
   Institute for Computational Cosmology, Department of Physics, Durham University, South Road, Durham DH1 3LE, UK
     \and
    GEPI, Observatoire de Paris, PSL Research University, CNRS, Place Jules Janssen, 92195 Meudon, France
    \and 
    Inter-University Centre for Astronomy and Astrophysics, Post Bag 4, Ganeshkhind, Pune, Maharashtra 411007, India
      }
   
   \date{Received XXX; accepted YYY}

% \abstract{}{}{}{}{} 
  \abstract
  {
   Boxy/peanut (b/p) bulges, the vertically extended inner parts of bars, are ubiquitous in barred galaxies in the local Universe, including our own Milky Way. At the same time, a majority of external galaxies and the Milky Way also possess a thick-disc. However, the dynamical effect of thick-discs in the b/p formation and evolution is not fully understood. Here, we investigate the effect of thick-discs in the formation and evolution of b/ps by using a suite of $N$-body models of (kinematically cold) thin and (kinematically hot) thick discs. Within the suite of models, we systematically vary the mass fraction of the thick disc, and the thin-to-thick disc scale length ratio. The b/ps form in almost all our models via a vertical buckling instability, even in the presence of a massive thick disc. The thin disc b/p is much stronger than the thick disc b/p. With increasing thick disc mass fraction, the final b/p structure gets progressively weaker in strength and larger in extent. Furthermore, the time-interval between the bar formation and the onset of buckling instability gets progressively shorter with increasing thick-disc mass fraction. The breaking and restoration of the vertical symmetry (during and after the b/p formation) show a spatial variation -- the inner bar region restores vertical symmetry rather quickly (after the buckling) while in the outer bar region, the vertical asymmetry persists long after the buckling happens. Our findings also predict that at higher redshifts, when discs are thought to be thicker, b/ps would have more `boxy-shaped' appearance than more `X-shaped' appearance. This remains to be tested from future observations at higher redshifts.
   }


   \maketitle
%
%-------------------------------------------------------------------

\section{Introduction}
\label{sec:Intro}
%&&&&&&&&&&&&&&&&&&&&&&&&&&&&&&&&&

A number of observational studies find that nearly half of all edge-on disc galaxies in the local Universe exhibit a prominent boxy or peanut-shaped structure \citep[hereafter b/p structure; e.g. see][]{BureauandFreeman1999,Luttickeetal2000,ErwinandDebattista2017}. A wide variety of observational and theoretical evidences indicate that many bars get vertically thickened in their inner regions, appearing as `boxy' or `peanut-shaped' bulges when seen in edge-on configuration \citep[e.g.][]{CombesandSanders1981,Rahaetal1991,ErwinandDebattista2016}. The presence of b/p structure has also been detected for galaxies in the face-on configurations, e.g. in IC~5240 \citep{Buta1995}, IC~4290 \citep{ButaandCrocker1991}, and several others \citep[see examples in][]{Quillenetal1997,McWilliamandZoccali2010,Laurikainenetal2011,ErwinandDebattista2013}. Several photometric and spectroscopic studies of the Milky Way bulge revealed that the Milky Way also has an inner b/p structure \citep[e.g., see][]{Natafetal2010,Shenetal2010,Nessetal2012,WeggandGerhard2013,Weggetal2015}. The occurrence of b/p bulges is observationally shown to depend strongly on the stellar mass of the galaxy, and a majority of barred galaxies above stellar mass $\log(M_*/M_{\odot}) \ge 10.4$ host b/p bulges \citep[e.g., see][]{YoshinoandYamauchi2015,ErwinandDebattista2017,Marchuketal2022}. A similar (strong) stellar mass dependence of the b/p bulge occurrence, at redshift $z =0$,  is shown to exist for the TNG50 suite of cosmological zoom-in simulation \citep{Andersonetal2023}. 
\par
Much of our current understanding of the b/p formation and its growth in barred galaxies are gleaned from numerical simulations. Studies using $N$-body simulations often find that, soon after the formation of a stellar bar it undergoes a vertical buckling instability, which subsequently gives rise to a prominent b/p bulge \citep[e.g., see][]{Combesetal1990,Rahaetal1991,MerrittandSellwood1994,Debattistaetal2004,Martinez-Valpuestaetal2006,Martinez-Valpuesta2008,Sahaetal2013}. Indeed, \citet{ErwinandDebattista2016} detected two such local barred-spiral galaxies which are undergoing such a buckling phase. If a barred $N$-body model is evolved for enough longer time, it might go through a second and prolonged buckling phase, thereby producing a prominent X-shape feature \citep[e.g.][]{Martinez-Valpuestaetal2006,Martinez-Valpuesta2008}. Furthermore, \citet{Sahaetal2013} showed that a bar buckling instability is closely linked with the maximum meridional tilt of the stellar velocity ellipsoid (denoting the meridional shear stress of stars).
Alternatively, a b/p bulge can be formed via the trapping of disc stars at vertical resonances during the secular growth of the bar \citep[e.g., see][]{CombesandSanders1981,Combesetal1990,Quillen2002,Debattistaetal2006,Quillenetal2014,Lietal2023} or by gradually increasing fraction of bar orbits trapped into this resonance \citep[e.g. see][]{SellwoodandGerhard2020}. The main difference between these two scenarios of b/p formation is that when the bar undergoes the buckling instability phase, the symmetry about the mid-plane is no longer preserved for a period of time \citep[e.g. see discussion in][]{Cuomoetal2022}. 
\par
Regardless of the formation scenario, the b/p bulges are shown to have a significant effect on the evolution of disc galaxies, by reducing the bar-driven gas inflow \citep[e.g. see][]{Fragoudietal2015,Fragoudietal2016,Athanassoula2016}. The formation of b/p bulges can affect metallicity gradients in the inner galaxy \citep[e.g.][]{Dimatteoetal2014,Fragkoudi2017}, and can also lead to bursts in star formation history \citep[e.g.][]{Perezetal2017}. In addition, \citet{Sahaetal2018} showed that for a 3-D b/p structure (i.e. b/p seen in both face-on and edge-on configurations), it introduces a kinematic pinch in the velocity map along the bar minor axis. Furthermore, \citet{Vynatheyaetal2021} demonstrated that for such a 3-D b/p structure, the inner bar region rotates slower than the outer bar region.
\par
On the other hand, a thick-disc component is now known to be ubiquitous in majority of external galaxies as well as in the Milky Way  \citep[e.g., see][]{Tsikoudi1979,Burstein1979,Yoachim2006,Comenronetal2011a,Comeron2011b,Comeronetal2018}. The existence of this thick-disc component covers the whole range of the Hubble classification scheme - from early-type, S0 galaxies to late-type galaxies \citep{Pohlenetal2004,Yoachim2006,Comeronetal2016,Kasparovaetal2016,Comeronetal2019,Pinnaetal2019b,Pinnaetal2019a,matigetal2021,Scottetal2021}. The thick-disc component is vertically more extended, and kinematically hotter as compared to the thin-disc component. The dynamical role of a thick-disc on the formation and growth of non-axisymmetric structures has been been studied for bars \citep[e.g.,][]{Klypinetal2009,AumerandBinney2017,Ghoshetal2022} and spirals \citep{GhoshJog2018,GhoshJog2021}. Past studies demonstrated that the (cold) thin and (hot) thick discs get mapped differently in the bar and boxy/peanut bulge \citep[e.g.][]{Athanassoulaetal2017,Fragkoudietal2017,Debattistaetal2017,Bucketal2019}. Since the presence of a thick disc can significantly affect the formation, evolution and properties of bars \citep{Ghoshetal2022}, we need to explore how it will affect the b/p's, since b/p's are essentially the vertical extended part of the bar.
\par
Stellar bars are known to be present in high redshift ($z \sim 1$) galaxies \citep[e.g., see][]{Shethetal2008,Elmetal2004,Jogeeetal2004,Guoetal2022,LeConteetal2023}. Furthermore, a recent study by \citet{Kruketal2019} showed the existence of b/p structure in high redshift ($z \sim 1$) galaxies as well. At high redshift, discs are known to be thick, kinematically hot (and turbulent), and more gas-rich. So, the question remains - how efficiently can the b/p structures form in such thick discs at such high redshifts?  \citet{Fragkoudietal2017} studied the effect of such a thick disc component on the b/p formation using a fiducial two-component thin+thick disc model where the thick-disc constitutes 30 percent of the total stellar mass. The formation and properties of b/p bulges in multi-component discs (i.e. with a number of disc populations greater than two) was studied also in \citet{DiMatteo2016}, \citet{Debattistaetal2017}, \citet{Fragkoudi2018}, \citet{Fragkoudietal2018}, and \citet{DiMatteoetal2019}. However, a systematic study of b/p formation in discs with different thin and thick discs, as well as composite thin and thick discs is still missing. We aim to pursue this here.
\par
In this paper, we systematically investigate the dynamical role of the thick-disc component in b/p formation and growth by using a suite of $N$-body models with (kinematically hot) thick and (kinematically cold) thin discs. We vary the thick-disc mass fraction as well as consider different geometric configurations (varying ratio of the thin- and thick-disc scale lengths) within the suite of $N$-body models. We quantify the strength and growth of the b/p in both the thin- and thick-disc stars, and also study the vertical asymmetry associated with the vertical buckling instability. In addition, we investigate the kinematic phenomena (i.e., change in the velocity dispersion, meridional tilt angle) associated with the b/p formation and its subsequent growth.
\par
The rest of the paper is organised as follows.
Sect.~\ref{sec:sim_setup} provides a brief description of the simulation set-up and the initial equilibrium models. 
Sect.~\ref{sec:bp_evolution_fthick} quantifies the properties of the b/p structure, their temporal evolution, as well as the vertical asymmetry in different models and the associated temporal evolution. Sect.~\ref{sec:kinematics} provides the details of kinematic phenomena related to the b/p formation and its growth while sect.~\ref{sec:Xshape_example} provides the details of relative contribution of thin-disc in supporting the X-shape structure. Sect.~\ref{sec:discussion} contains the discussion while  Sect.~\ref{sec:conclusion} summarizes the main findings of this work.

\section{Simulation set-up \& $N$-body models}
\label{sec:sim_setup}
%&&&&&&&&&&&&&&&&&&&&&&&&&&&&&&&&&

\begin{table*}
\centering
\caption{Key structural parameters for the equilibrium models.}
\begin{tabular}{cccccccccccc}
\hline
\hline
 Model$^{(1)}$ & $f_{\rm thick}$$^{(2)}$ & $R_{\rm d, thin}$$^{(3)}$ & $R_{\rm d, thick}$$^{(4)}$ & &  $z_{d, \rm {thin}}$$^{(5)}$ & $z_{d, \rm {thick}}$$^{(6)}$ & $M_{\rm star}$$^{(7)}$ & $R_{\rm H}$$^{(8)}$ & $M_{\rm dm}$$^{(9)}$ & $n_{\rm star}$$^{(10)}$ & $n_{\rm dm}$$^{(11)}$ \\
\\
 && (kpc) & (kpc) & & (kpc) & (kpc) & ($\times 10^{11} M_{\odot}$) & (kpc) & ($\times 10^{11} M_{\odot}$) & ($\times 10^{5}$) & ($\times 10^{5}$)  \\
\hline
rthick0.0 & -- & 4.7 &  -- &  & 0.3 & -- & 1 & 10 & 1.6 & 10 & 5 \\
\hline
rthickS0.1 & 0.1 & 4.7 &  2.3 &  & 0.3 & 0.9 & 1 & 10 & 1.6 & 10 & 5 \\
rthickE0.1 & 0.1 & 4.7 &  4.7 &  & 0.3 & 0.9 & 1 & 10 & 1.6 & 10 & 5 \\
rthickG0.1 & 0.1 & 4.7 &  5.6 &  & 0.3 & 0.9 & 1 & 10 & 1.6 & 10 & 5 \\
rthickS0.3 & 0.3 & 4.7 &  2.3 &  & 0.3 & 0.9 & 1 & 10 & 1.6 & 10 & 5 \\
rthickE0.3 & 0.3 & 4.7 &  4.7 &  & 0.3 & 0.9 & 1 & 10 & 1.6 & 10 & 5 \\
rthickG0.3 & 0.3 & 4.7 &  5.6 &  & 0.3 & 0.9 & 1 & 10 & 1.6 & 10 & 5 \\
rthickS0.5 & 0.5 & 4.7 &  2.3 &  & 0.3 & 0.9 & 1 & 10 & 1.6 & 10 & 5 \\
rthickE0.5 & 0.5 & 4.7 &  4.7 &  & 0.3 & 0.9 & 1 & 10 & 1.6 & 10 & 5 \\
rthickG0.5 & 0.5 & 4.7 &  5.6 &  & 0.3 & 0.9 & 1 & 10 & 1.6 & 10 & 5 \\
rthickS0.7 & 0.7 & 4.7 &  2.3 &  & 0.3 & 0.9 & 1 & 10 & 1.6 & 10 & 5 \\
rthickE0.7 & 0.7 & 4.7 &  4.7 &  & 0.3 & 0.9 & 1 & 10 & 1.6 & 10 & 5 \\
rthickG0.7 & 0.7 & 4.7 &  5.6 &  & 0.3 & 0.9 & 1 & 10 & 1.6 & 10 & 5 \\
rthickS0.9 & 0.9 & 4.7 &  2.3 &  & 0.3 & 0.9 & 1 & 10 & 1.6 & 10 & 5 \\
rthickE0.9 & 0.9 & 4.7 &  4.7 &  & 0.3 & 0.9 & 1 & 10 & 1.6 & 10 & 5 \\
rthickG0.9 & 0.9 & 4.7 &  5.6 &  & 0.3 & 0.9 & 1 & 10 & 1.6 & 10 & 5 \\
rthickS1.0 & 1 & -- &  2.3 &  & -- & 0.9 & 1 & 10 & 1.6 & 10 & 5 \\
rthickE1.0 & 1 & -- &  4.7 &  & -- & 0.9 & 1 & 10 & 1.6 & 10 & 5 \\
rthickG1.0 & 1 & -- &  5.6 &  & -- & 0.9 & 1 & 10 & 1.6 & 10 & 5 \\
\hline
\end{tabular}
\newline
{\newline
\textbf{Notes. } Column (1): Name of the model. Col. (2): Thick-disc mass fraction. Col. (3): Scale length of the thin disc. Col. (4): Scale length of the thick disc. Col. (5): Scale height of the thin disc. Col. (6): Scale height of the thick disc. Col. (7): Mass of the stellar (thin+thick) disc. Col. (8): Characteristic scale length of the dark matter halo. Col. (9): Mass of the dark matter halo. Col. (10): Total number of particles in the stellar (thin+thick) disc. Col. (11): Total number of particles in the dark matter halo. }
\label{table:key_param}
\end{table*}
To motivate our study, we use a suite of $N$-body models, consisting of a thin and a thick stellar disc, and the whole system is embedded in a live dark matter halo. One such model is already presented in \citet{Fragkoudietal2017}.
In addition, these models have been thoroughly studied in a recent work of \citet{Ghoshetal2022} in connection with bar formation scenario under varying thick-disc mass fractions. Here, we use the same suite of thin+thick models to investigate b/p formation and evolution with varying thick-disc mass fraction.
\par
The details of the initial equilibrium models and how they are generated are already given in \citet{Fragkoudietal2017} and \citet{Ghoshetal2022}. Here, for the sake of completeness, we briefly mention the equilibrium models. Each of the thin- and thick-discs is modelled with a Miyamoto-Nagai profile \citep{MiyamatoandNagai1975}, having $R_{\rm d}$, $z_{\rm d}$, and $M_{\rm d}$ as the characteristic disc scale length, the scale height, and the total mass of the disc, respectively. The dark matter halo is modelled with a Plummer sphere \citep{Plummer1911}, having $R_{\rm H}$ and $M_{\rm dm}$ as the characteristic scale length and the total halo mass, respectively. The values of the key structural parameters for the thin- and thick-discs as well as the dark matter halo, and the total number of particles used to model each of these structural components are mentioned in Table.~\ref{table:key_param}. For this work, we analysed a total of 19 $N$-body models (including one pure thin-disc-only and 3 pure thick-disc-only models) of such thin+thick discs.
\par
The initial conditions of the discs are obtained using the iterative method algorithm \citep[see][]{Rodionovetal2009}. For further details, the reader is referred to \citet{Fragkoudietal2017} and \citet{Ghoshetal2022}. The simulations are run using a TreeSPH code by \citet{SemelinandCombes2002}. A hierarchical tree method \citep{BarnesandHut1986} with opening angle $\theta = 0.7$ is used for calculating the gravitational force which includes terms up to the quadrupole order in the multipole expansion. A Plummer potential is employed for softening the gravitational forces with a softening length $\epsilon = 150 \pc$. We evolved all the models for a total time of $9 \Gyr$.
\par
Within the suite of thin+thick disc models, we consider three different scenarios for the scale lengths of the two disc (thin and thick) components. In rthickE models, both the scale lengths of thin- and thick-discs are kept same ($R_{\rm d, thick} = R_{\rm d, thin}$). In rthickS models, the scale length of the thick-disc component is shorter than that for the thin-disc ($R_{\rm d, thick} < R_{\rm d, thin}$), and in rthickG models, the scale length of the thick-disc component is larger than that for the thin-disc ($R_{\rm d, thick} > R_{\rm d, thin}$). 
Following the nomenclature scheme used in \citet{Ghoshetal2022}, any thin+thick model is referred as a unique string `{\sc [model configuration][thick disc fraction]'}. {\sc [model configuration]} denotes the corresponding thin-to-thick disc scale length configuration, that is, rthickG, rthickE, or {rthickS whereas {\sc [thick disc fraction]} denotes the fraction of the total disc stars that are in the thick-disc population (or equivalently the mass fraction in the thick-disc as all the disc particles have same mass).
\par
Before we present the results, we mention that in our thin+thick models, we can identify and separate, by construction, which stars are members of the thin-disc component at initial time ($t = 0$) and which stars are members of the thick-disc component at $t =0$, and we can track them as the system evolves self-consistently. Thus, throughout this paper, we refer to the b/p as seen exclusively in particles initially belonging to the thin-disc population as the `thin disc b/p' and that seen exclusively in particles initially belonging to the thick-disc population as the `thick disc b/p'.

\section{Boxy/peanut formation \& evolution for different mass fraction of thick disc population}
\label{sec:bp_evolution_fthick}
%&&&&&&&&&&&&&&&&&&&&&&&&&&&&&&&&&

%&&&&&&&&&&&&&&&&&&&&&&
\subsection{Quantifying the b/p properties}
\label{sec:bp_properties}
%&&&&&&&&&&&&&&&&&&&&&&

%##################
% Begin figure
%##################
\begin{figure*}
\centering
\resizebox{\linewidth}{!}{\includegraphics{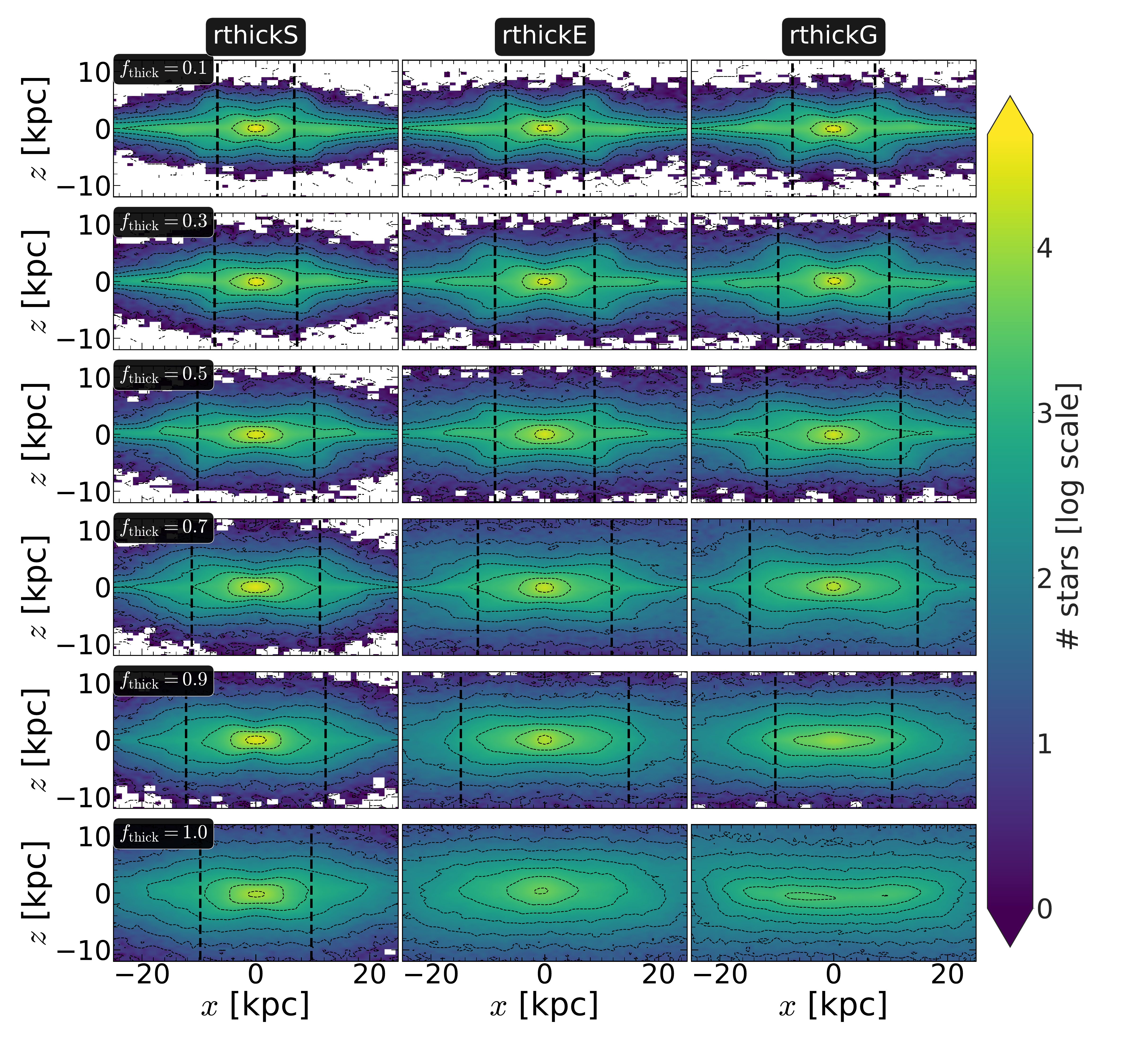}}
\caption{Edge-on density distribution of all disc particles (thin+thick), at the end of the simulation run ($t = 9 \Gyr$) for all thin+thick disc models with varying $f_{\rm thick}$ values. Black dotted lines denote the contours of constant density.  \textit{Left panels} show the density distribution for the rthickS models whereas  \textit{middle panels} and \textit{right panels} show the density distribution for the rthickE  and rthickG models, respectively. The thick disc fraction ($f_{\rm thick}$) varies from 0.1 to 1 (top to bottom), as indicated in the left-most panel of each row. The bar is placed along the $x$-axis (side-on configuration) for each model. The vertical black dashed lines denote the extent of the b/p structure in each case, for details see the text in Sect.~\ref{sec:bp_length_temporal_evolution}.}
\label{fig:density_maps_endstep_allmodels}
\end{figure*}
%##################
% End figure
%##################

%##################
% Begin figure
%##################
\begin{figure*}
\centering
\resizebox{0.95\linewidth}{!}{\includegraphics{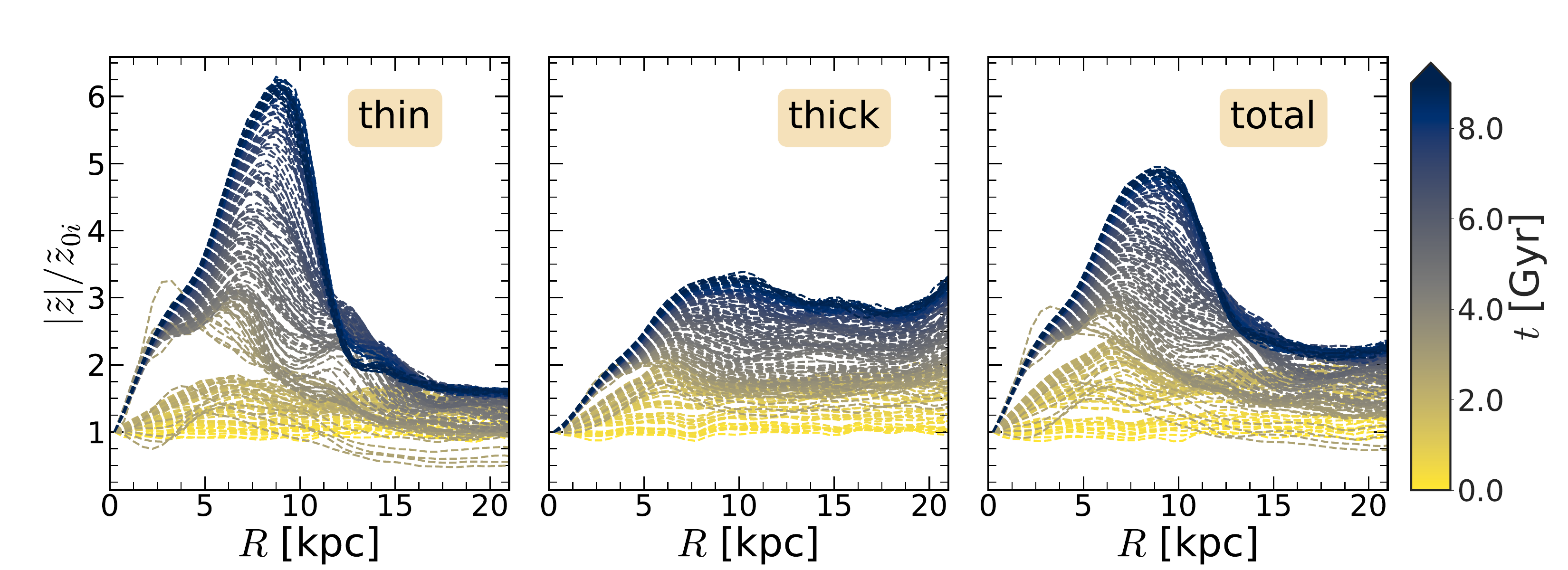}}
\caption{Radial profiles of the median of the absolute value of the distribution of particles in the vertical ($z$) direction, $|\tilde z|$, (normalised by the initial value, $\tilde z_0$), for the thin-disc (left panels), thick-disc stars (middle panels), and total (thin+thick) disc stars (right panels) as a function of time (shown in colour bar) for the model rthickE0.5.  Here, $\tilde z_{0i}$ denotes the initial value (used for the normalization, for details see the text) where $i=$ thin, thick, thin+thick, respectively. The thin disc b/p remains much stronger as compared to the thick disc b/p.}
\label{fig:radialBPstrength_profiles_rthickE05}
\end{figure*}
%##################
% End figure
%##################

%##################
% Begin figure
%##################
\begin{figure*}
\centering
\resizebox{0.95\linewidth}{!}{\includegraphics{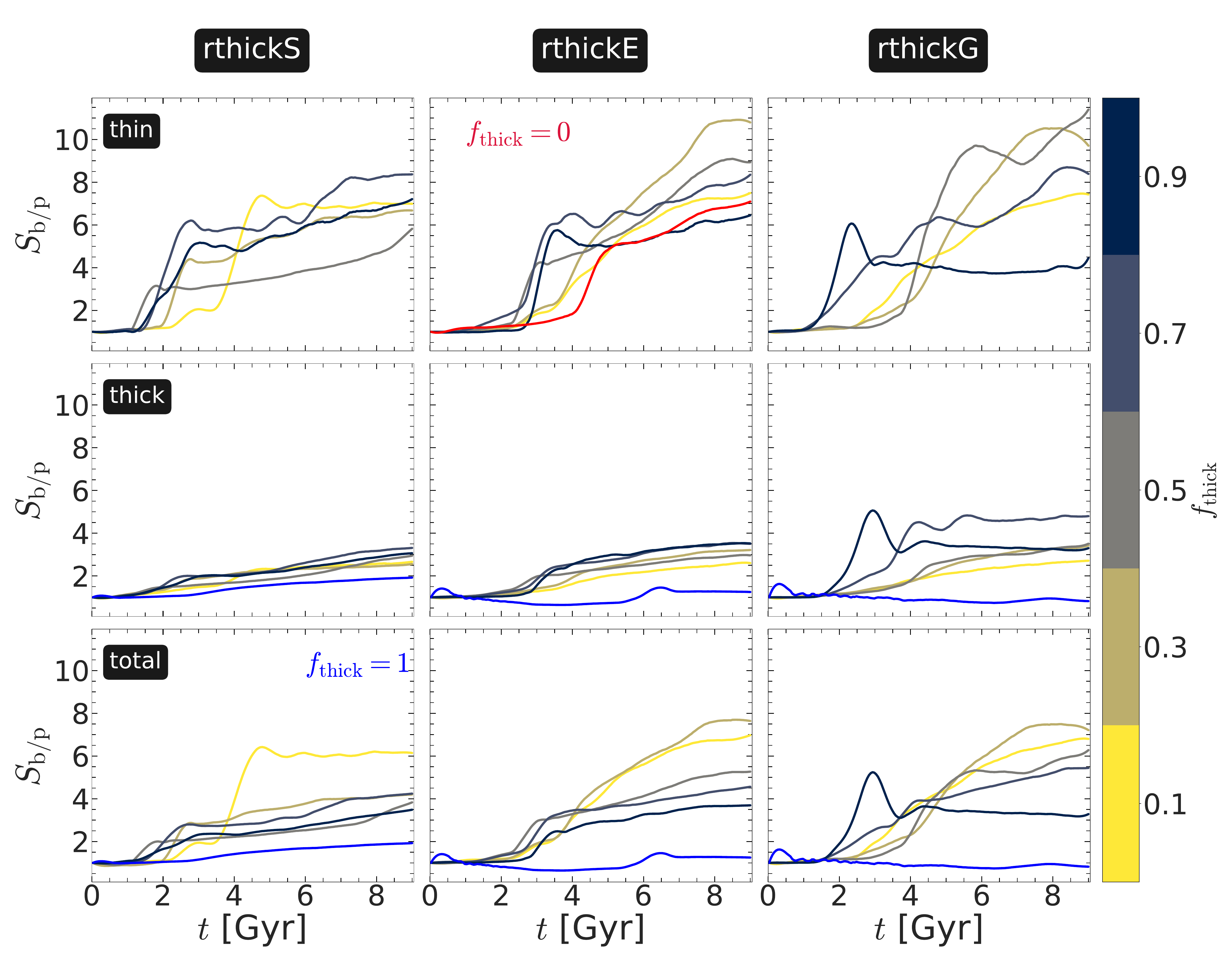}}
\caption{Temporal evolution of the b/p strength, $S_{\rm b/p}$, calculated using Eq.~\ref{eq:bp_strength}, for the thin-disc (upper panels), thick-disc (middle panels), and total (thin+thick) disc stars (lower panels) for all thin+thick disc models with varying $f_{\rm thick}$ values (see the colour bar). \textit{Left panels} show the b/p strength evolution for the rthickS models whereas  \textit{middle panels} and \textit{right panels} show the b/p strength evolution for the rthickE  and rthickG models, respectively. The thick disc fraction ($f_{\rm thick}$) varies from 0.1 to 0.9 (with a step-size of 0.2), as indicated in the colour bar. The blue solid lines in the middle and the bottom rows denote the three thick disc-only models ($f_{\rm thick} =1$) whereas the red solid line in the top middle panel denotes the thin disc-only model ($f_{\rm thick} =0$). For details, see the text.}
\label{fig:bpStrength_evolution}
\end{figure*}
%##################
% End figure
%##################

Fig.~\ref{fig:density_maps_endstep_allmodels} shows the distribution of all stars (thin+thick) in the edge-on projection, calculated at the end of the simulation run ($t = 9 \Gyr$), for all the thin+thick models considered here. In each case, the bar is placed in the side-on configuration (i.e., along the $x$-axis). A prominent b/p structure is seen in most of these thin+thick models. We further checked the same edge-on stellar density distribution, separately calculated for the thin- and thick-disc stars. Both of them show a prominent b/p structure in most of the thin+thick models. For the sake of brevity, we have not shown it here \citep[however, see Fig.~2 in][]{Fragkoudietal2017}. 

\subsubsection{Quantifying the b/p strength \& its temporal evolution}
\label{sec:bp_stregth_temporal_evolution}
%&&&&&&&&&&&&&&&&&&&&&&&&&&&&&

Here, we quantify the strength of the b/p structure and study its variation (if any) with thick-disc mass fraction. Following \citet{Martinez-Valpuesta2008} and \citet{Fragkoudietal2017}, in a given radial bin of size $\Delta R \ (= 0.5 \kpc)$, we calculate the median of
the absolute value of the distribution of particles in the vertical ($z$) direction, $|\tilde z|$, (normalised by the initial value, $\tilde z_0$) of a snapshot seen edge-on, and with the bar placed side-on (along the $x$-axis). 
In Fig.~\ref{fig:radialBPstrength_profiles_rthickE05}, we show one example of the  corresponding radial profiles of the $|\tilde z|/ \tilde z_{0i}$ ($i=$ thin, thick, thin+thick), computed separately for the thin- and thick-disc particles, as well as for the thin+thick disc particles, as a function of time. As seen in Fig~\ref{fig:radialBPstrength_profiles_rthickE05}, a prominent peak in the radial profiles (at later times) of $|\tilde z|/ \tilde z_{0i}$, denotes the formation of a b/p structure in the thin+thick model. Here, we mention that as the vertical scale height, and hence the vertical extent of thick disc is larger (by a factor of 3) than that for the thin disc (see Table.~\ref{table:key_param}), the normalisation by $\tilde z_0$ is necessary to unveil the intrinsic vertical growth due to the b/p formation. When only the absolute values of $|\tilde z|$ are considered, the thick-disc stars always produce a larger value of $|\tilde z|$ than the thin-disc stars. This happens due to the construction of equilibrium thin+thick models, and not due to the b/p formation.
The normalised peak for the thin-disc is much larger than that for the thick-disc, in concordance with the previous results \citep{Fragkoudietal2017}. Furthermore, at later times, the peak in $|\tilde z|/ \tilde z_{0i}$ profiles shifts towards outer radial extent (more prominent for the thin disc b/p), indicating the growth of the b/p structure  towards outer radial extent. These trends are seen to hold for other thin+thick models as well which form a b/p structure during their evolutionary pathway.
\par
To quantify the temporal evolution of the b/p strength, we define the b/p strength at time $t$, $S_{\rm b/p} (t)$ as the maximum of the peak value of the  $|\tilde z|/ \tilde z_{0i}$, i.e.,
\begin{equation}
S_{\rm b/p} (t) = {\rm max} \left( \frac{|\tilde z|}{\tilde z_{0i}} \right)\,.
\label{eq:bp_strength}
\end{equation}
In \citet{Martinez-Valpuestaetal2006} and \citet{Martinez-Valpuesta2008}, a method based on the Fourier decomposition has been formulated to quantify the strength of a b/p structure. In Appendix~\ref{appen:bp_strength_fourier}, we compare this method with Eq.~\ref{eq:bp_strength} for the thin+thick model rthickE0.5.
\par

 Fig.~\ref{fig:bpStrength_evolution} shows the corresponding temporal evolution of the b/p strength, calculated separately for the thin- and thick-discs, as well as for the composite thin+thick disc particles, for all thin+thick models considered here. The thin disc b/p remains much stronger than the thick disc b/p structure, and this trend holds true for all thin+thick models with three different configurations (i.e., rthickS, rthickE, and rthickG) which develop a prominent b/p structure during their course of temporal evolution. 

\par
The temporal evolution of the b/p strength in three thick-disc-only models merit some discussion. For the rthickS1.0 model, the values of $S_{\rm b/p}$ show a monotonic increase with time, denoting the formation of a prominent b/p structure (also see the bottom row of Fig.~\ref{fig:density_maps_endstep_allmodels}). However, the final b/p strength for the rthickS1.0 model remains lowest when compared to other rthickS models with different $f_{\rm thick}$ values. For the rthickE1.0 model, the temporal evolution of $S_{\rm b/p}$ shows a sudden jump around $t = 6.5 \Gyr$ and then remains constant. By the end of the simulation, this model forms a b/p structure which appears more `boxy' rather than a `peanut/X-shape'. For the rthickG1.0 model, the temporal evolution of $S_{\rm b/p}$ does not show much increment, and the model does not form a prominent b/p structure at the end of the simulation.
 
%##################
% Begin figure
%##################
\begin{figure}
\centering
\resizebox{\linewidth}{!}{\includegraphics{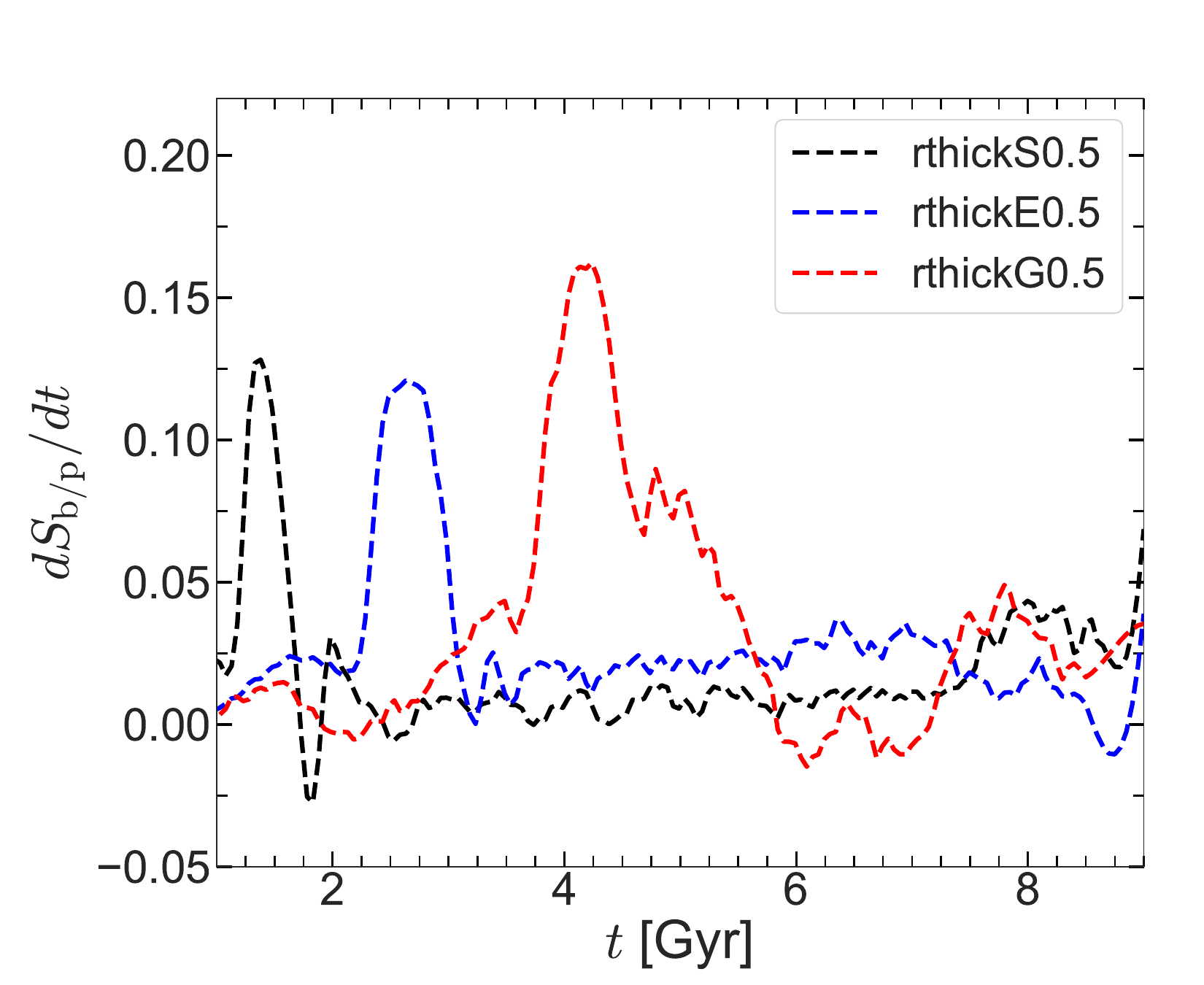}}
\caption{The growth rate of the b/p, $d S_{\rm b/p}/dt$, as a function of time for  three thin+thick models with $f_{\rm thick} = 0.5$. All the stellar particles (thin+thick) are considered here in each model. The growth rate is steeper in rthickS0.5 model when compared with that for other two configurations with same $f_{\rm thick}$ value.}
\label{fig:bp_growthrate_rthick05}
\end{figure}
%##################
% End figure
%##################
 
In addition, for a fixed value of $f_{\rm thick}$, we calculated the gradient of $S_{\rm b/p} (t)$ with respect to time $t$ ($d S_{\rm b/p} (t)/dt $) for three different geometric configurations considered here. One such example is shown in Fig.~\ref{fig:bp_growthrate_rthick05} for $f_{\rm thick} = 0.5$. A prominent (positive) peak in the  $d S_{\rm b/p}/dt$ profile denotes the onset of the b/p formation. As seen clearly, for a fixed $f_{\rm thick}$ value, the peak in the $d S_{\rm b/p}/dt$ profile occurs at an earlier epoch when compared with the other two geometric configurations. This confirms that the b/p forms at an earlier time in rthickS0.5 model as compared to the rthickE0.5, and rthickG0.5 models. These trends are in tandem with the fact that the bar in rthickS models form at earlier times and grows faster as compared to the other two disc configurations \citep[for details, see][]{Ghoshetal2022}.
 \par
 Lastly, in Fig.~\ref{fig:bplength_evolution_endstep} (top panel), we show the final b/p strength (i.e., calculated at $t = 9 \Gyr$ using the thin+thick disc particles) for all thin+thick models considered here. We point out that, in some cases, the maximum of the $|\tilde z|/ \tilde z_{0i}$ profile for the thick-disc is not always easy to locate; sometimes they display a plateau rather than a clear maximum (see Fig.~\ref{fig:radialBPstrength_profiles_rthickE05}). This, in turn, might have an impact in the estimate of the b/p strength. In order to derive an estimate of the uncertainty on the b/p strength measurement, we constructed a total of 5,000 realisations by resampling the entire population via bootstrapping technique \citep{Pressetal1986}, and for each realisation, we computed the radial profiles of $|\tilde z|/ \tilde z_{0i}$  as well as its peak value (denoting the b/p strength). The resulting error estimates are shown in Fig.~\ref{fig:bplength_evolution_endstep}. The final b/p strength shows a wide variation with the $f_{\rm thick}$ values as well as with the thin-thick disc configuration. To illustrate, for the rthickS models, the final b/p strength decreases monotonically as  $f_{\rm thick}$ value increases. For the rthickE models, the final b/p strength increases from $f_{\rm thick} = 0.1 - 0.3$, and then decreases monotonically as  $f_{\rm thick}$ value increases. A similar trend is also seen for the rthickG models. Nevertheless, the strength of the b/p shows an overall decreasing trend with increasing $f_{\rm thick}$ values, and this remains true for all three geometric configurations considered here.

%##################
% Begin figure
%##################
\begin{figure}
\centering
\resizebox{\linewidth}{!}{\includegraphics{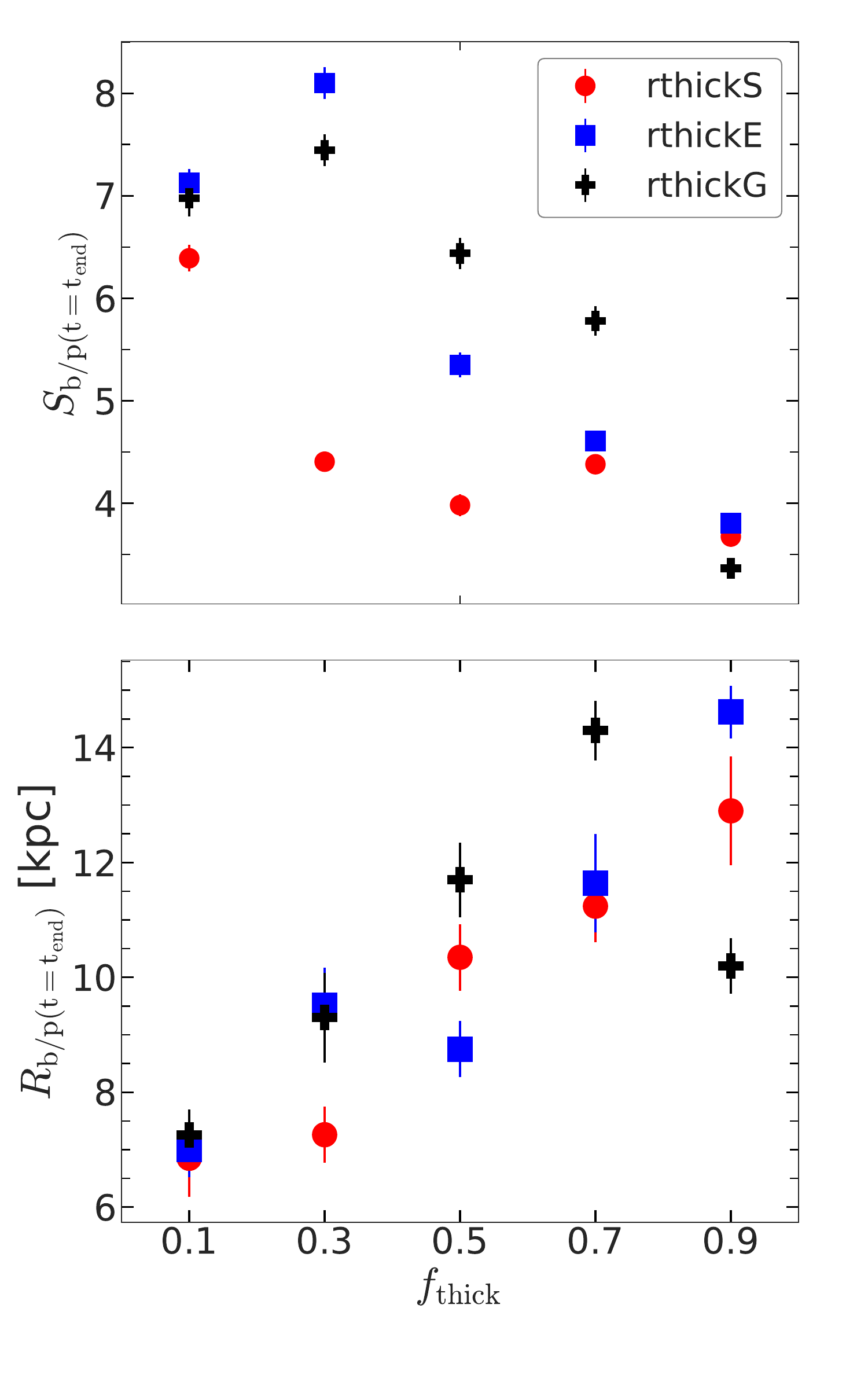}}
\caption{The strength of the b/p, $S_{\rm b/p}$ (\textit{top panel}), and the extent of the b/p, $R_{\rm b/p}$ (\textit{bottom panel}), calculated using thin+thick disc particles, at $t = 9 \Gyr$ shown as a function of thick-to-total mass fraction ($f_{\rm thick}$) and for different geometric configurations. With increasing $f_{\rm thick}$ value, the b/ps progressively become weaker and larger in extent, and this trend remains true for all three geometric configurations considered here. The errors are calculated by constructing a total of 5,000 realisations by resampling the entire population via bootstrapping technique. For details, see the text.} 
\label{fig:bplength_evolution_endstep}
\end{figure}
%##################
% End figure
%##################

\subsubsection{Quantifying the b/p length \& its temporal evolution}
\label{sec:bp_length_temporal_evolution}
%&&&&&&&&&&&&&&&&&&&&&&&&&&&&&

The extent of the b/p structure is another quantity of interest, and it is worth studying whether the b/p extent vary across different disc configurations and $f_{\rm thick}$ values. At time $t$, we define the extent of the b/p structure, $R_{\rm b/p}$, as the radial location where the peak in $|\tilde z|/ \tilde z_{0i}$ profile occurs. In \citet{Luttickeetal2000} and \citet{SahaandGerhard2013}, a method based on the line-of-sight surface density profile has been formulated to quantify the size of a b/p structure. In Appendix~\ref{appen:bp_length_fourier}, we compare this method with $R_{\rm b/p}$ for the thin+thick model rthickE0.5.
In Fig.~\ref{fig:bar_strength_rthickE05}, we show the temporal evolution of $R_{\rm b/p}$ for the thin+thick model rthickE0.5. The errors on the b/p length are estimated using the same bootstrapping technique mentioned in Sect.~\ref{sec:bp_stregth_temporal_evolution}. The b/p length increases significantly (by factor of $\sim 2$) over the entire evolutionary phase. In addition, towards the end of the simulation run, the thick disc b/p is larger (by $\sim 10-15$ percent) than the thin disc b/p. We found a similar trend in temporal variation of the b/p extent for other thin+thick models, and therefore, they are not shown here.
\par
We further checked how the extent of the b/p structure, by the end of the simulation run, vary with the thin-to-thick disc mass fraction ($f_{\rm thick}$). In Fig.~\ref{fig:bplength_evolution_endstep} (bottom panel), we show the corresponding extents of the b/p structure, computed at $t = 9 \Gyr$, using the thin+thick stellar particles, for all thin+thick models considered here. The extent of the b/p structure increases steadily as $f_{\rm thick}$ value increases, and this trend holds true for all three different configurations considered here. Furthermore, at a fixed $f_{\rm thick}$ value, the rthickG models show a higher value for the $R_{\rm b/p}$ when compared to other two configurations, thereby denoting that rthickG models form a larger b/p structure, by the end of the simulation run, as compared to rthickE and rthickS models. Lastly, in Appendix~\ref{appen:bp_length_componentwise}, we show how the extent of the thin disc b/p and thick disc b/p, at the end of the simulation ($t = 9 \Gyr$), vary across different $f_{\rm thick}$ values and different disc configurations. 

%##################
% Begin figure
%##################
\begin{figure}
\centering
\resizebox{1.0\linewidth}{!}{\includegraphics{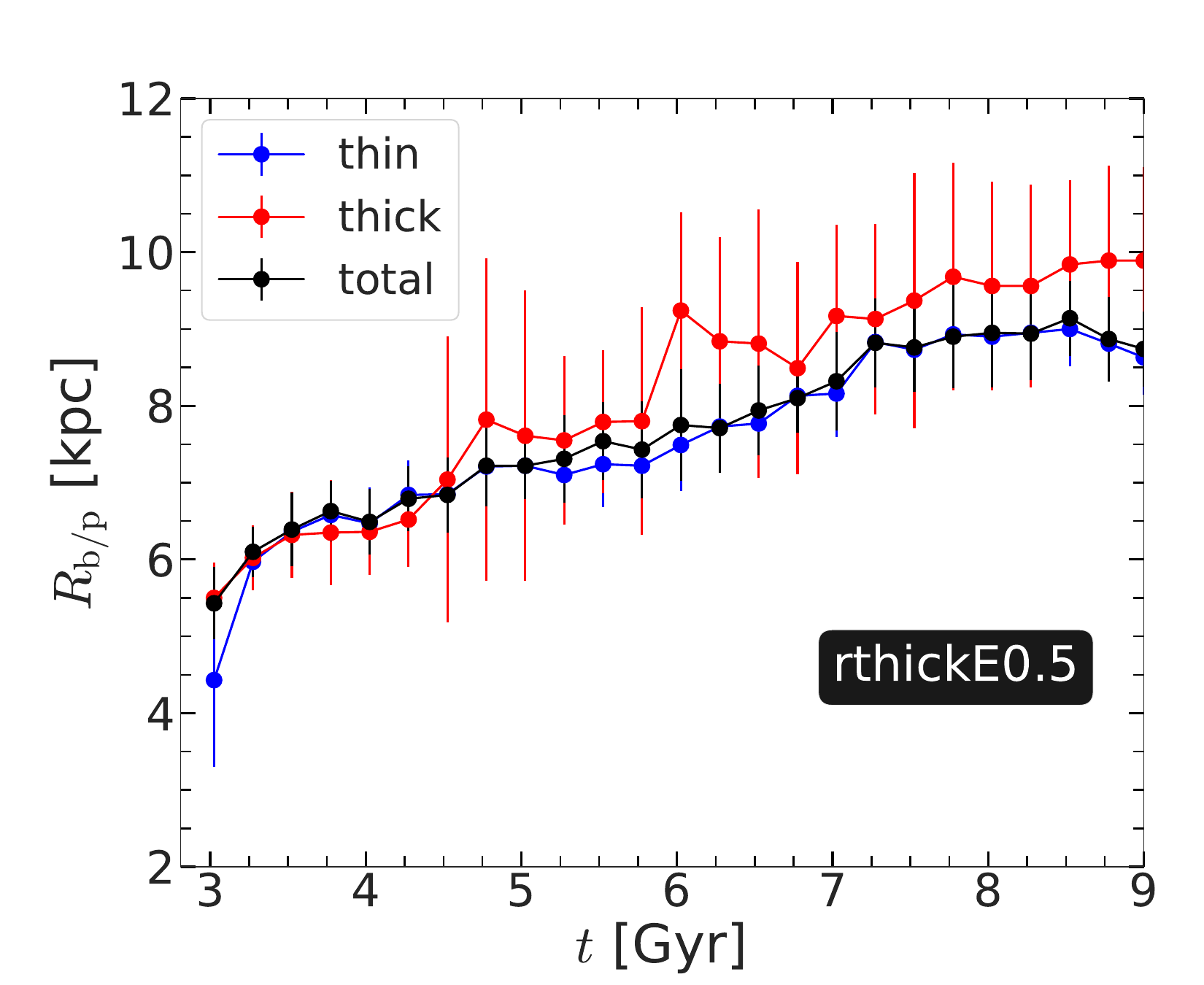}}
\caption{Temporal variation of the b/p extent, $R_{\rm b/p}$, calculated using thin-, thick-disc, and thin+thick disc particles for the model rthickE0.5. The b/p extent increases by a factor of $\sim 2$ over the total simulation runtime. At $t = 9 \Gyr$, the thick disc b/p remains a bit larger than the thin disc b/p.  The errors on $R_{\rm b/p}$ are estimated by constructing a total of 5,000 realisations by resampling the entire population via bootstrapping technique. For details, see the text.}
\label{fig:bar_strength_rthickE05}
\end{figure}
%##################
% End figure
%##################

%&&&&&&&&&&&&&&&&&&&&&&&&&&
\subsection{Vertical asymmetry and buckling instability}
\label{sec:vert_asymmetry}

%&&&&&&&&&&&&&&&&&&&&&&&&&&
%##################
% Begin figure
%##################
\begin{figure*}
\centering
\resizebox{0.85\linewidth}{!}{\includegraphics{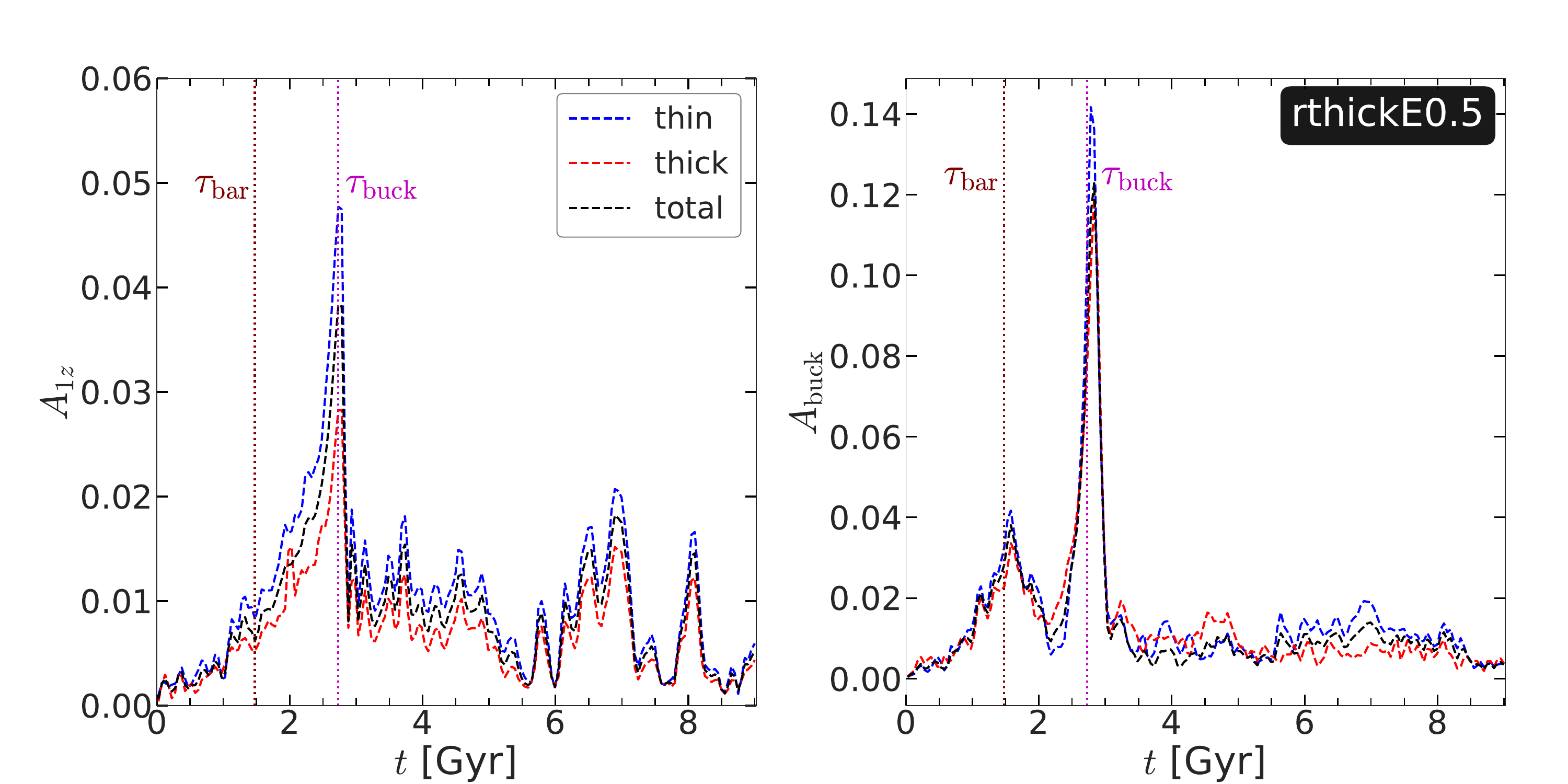}} 
\caption{\textit{Left panel :} temporal evolution of $A_{1,z}$ (Eq.~\ref{eq:fourier_buckling}) denoting the vertical asymmetry in the bar region, calculated using thin-disc (blue lines), thick-disc (red lines), and thin+thick disc (black lines), for the model rthickE0.5. \textit{Right panel :} temporal evolution of the buckling amplitude, $A_{\rm buck}$ (Eq.~\ref{eq:buckling_amp}), calculated using thin-disc (blue lines), thick-disc (red lines), and thin+thick disc (black lines), for the model rthickE0.5. The vertical magenta dotted line denotes the onset of buckling instability ($\tau_{\rm buck}$), calculated from the peak of $A_{\rm buck}$ profile. For details, see the text. Furthermore, for reference, we indicate the onset of bar formation ($\tau_{\rm bar}$, vertical maroon dotted line), calculated from the amplitude of the $m=2$ Fourier moment. }
\label{fig:temporal_A1z_allmodels}
\end{figure*}
%##################
% End figure
%##################

To quantify the vertical asymmetry and the time at which it occurs, we first calculate the amplitude of the first coefficient in the Fourier
decomposition ($A_{1z}$) which provides a measure of the asymmetry \citep[e.g. see][]{Martinez-Valpuestaetal2006,Martinez-Valpuesta2008,Sahaetal2013}. The first Fourier coefficient ($A_{1z}$) is defined as \citep{Martinez-Valpuesta2008}
\begin{equation}
A_{1z} = \frac{1}{\pi \ M_{\rm tot}} \left| \sum_{i} m_i e^{i m \varphi_i}\right|; \ m=1\,,
\label{eq:fourier_buckling}
\end{equation}
\noindent where $m_i$ is the mass of the $i$th particle, and $\varphi_i$ is the angle of $i$th particle measured in the $(x, z)-$plane with the bar placed along the $x$-axis (side-on configuration). $M_{\rm tot}$ is denotes the total mass of the particles considered in this summation. Following \citet{Martinez-Valpuesta2008}, to make this coefficient more sensitive to a buckling, we only included the stars in the summation (see Eq.~\ref{eq:fourier_buckling}) that are momentarily within the extent of the b/p ($R_{\rm b/p}$). The corresponding temporal evolution of the buckling amplitude ($A_{1z}$), calculated separately for thin, thick, and thin+thick particles, for the model rthickE0.5 is shown in Fig.~\ref{fig:temporal_A1z_allmodels} (left panel). A prominent peak in the $A_{1z}$ profile denotes the vertical buckling event. We further checked that for all the thin+thick models, a peak in the $A_{1z}$ is associated with the dip/decrease in the bar strength. This is expected since it is well known that the bar strength decreases as it goes through the buckling phase. The $A_{1z}$ amplitude is larger for the thin-disc stars when compared with that of the thick-disc stars. This is consistent is with the scenario that the thin disc b/p is stronger than the thick disc b/p.
\par
Another way of quantifying the buckling instability is via measuring the buckling amplitude, $A_{\rm buck}$ which is defined as \citep[for details, see][]{Sellwood1986,Debattistaetal2006,Debattistaetal2020}
\begin{equation}
A_{\rm buck} = \left|\frac{\sum_j z_j m_j e^{2i \phi_j} }{\sum_j m_j} \right|\,,
\label{eq:buckling_amp}
\end{equation}
\noindent where $m_j$, $z_j$, and $\phi_j$ denote the mass, vertical position, and azimuthal angle of the $j$th particle, respectively, and the summation runs over all star particles (thin, thick, thin+thick, whichever is applicable) within the b/p extent. The quantity $A_{\rm buck}$ denotes the $m =2$ vertical bending amplitude \citep[for further details, see][]{Debattistaetal2006,Debattistaetal2020}. The corresponding temporal evolution of the buckling amplitude ($A_{\rm buck}$), calculated separately for thin, thick, and thin+thick particles,  for the model rthickE0.5 is shown in Fig.~\ref{fig:temporal_A1z_allmodels} (right panel). A prominent peak in the $A_{\rm buck}$ profile denotes the onset of the vertical buckling instability. Furthermore, the peak of value of $A_{\rm buck}$ is higher for the thin-disc  than that for the thick-disc. This is again consistent with the thin disc b/p being stronger than the thick disc b/p. This trend holds true for all the thin+thick models considered here.
Next, we define $\tau_{\rm buck}$ as the epoch for onset of the buckling event when the peak in $A_{\rm buck}$ occurs.  As seen from Fig.~\ref{fig:temporal_A1z_allmodels}, the epoch at which the peak in $A_{1z}$ occurs, coincides with $\tau_{\rm buck}$. This is not surprising as both the quantities denote the same physical phenomenon of vertical buckling instability. In Sect.~\ref{sec:bp_bar_correlation}, we further investigate the variation of $\tau_{\rm buck}$ with $f_{\rm thick}$, and its connection with bar formation epoch.

\par
While Fig.~\ref{fig:temporal_A1z_allmodels} clearly demonstrates the temporal evolution of the vertical asymmetry associated with the b/p structure formation, it should be borne in mind that $A_{1z}$ (quantifying vertical asymmetry) or $A_{\rm buck}$ (quantifying the $m =2$ vertical bending amplitude) only informs us about the buckling instability in some average sense, and hence lacks any information about the two-dimensional distribution of the vertical asymmetry. To investigate that we computed the two-dimensional distribution of the mid-plane asymmetry. Following \citet{Cuomoetal2022}, we define 
\begin{equation}
A_{\Sigma} (x, z) = \frac{\Sigma(x, z) - \Sigma(x, -z)}{\Sigma(x, z) + \Sigma(x, -z)}\,,
\label{eq:midplane_asymmetry}
\end{equation}
\noindent where $\Sigma(x, z)$ denotes the projected surface number density of the particles at each position of the image of the edge-on view of the model. The resulting distribution of $A_{\Sigma} (x, z)$, computed separately for the thin-disc, thick-disc, and thin+thick discs at six different times (before and after the buckling happens) are shown in Fig.~\ref{fig:temporal_verticalasymmetry_anothermethod} for the model rthickE0.5. At the initial rapid bar growth phase ($t \sim 1 \Gyr$), the $A_{\Sigma} (x, z)$ values remain close to zero, indicating no breaking of vertical symmetry in that evolutionary phase of the model. Around $t \sim 2.7 \Gyr$, the model undergoes a strong buckling event (see the peak in Fig.~\ref{fig:temporal_A1z_allmodels}). As a result, the distribution $A_{\Sigma} (x, z)$ shows large positive/negative values at $t = 2.75 \Gyr$, thereby demonstrating that the vertical symmetry is broken about the mid-plane. At a later time ($t = 5 \Gyr$), the vertical symmetry is restored in the inner region (as indicated by $A_{\Sigma} (x, z) \sim 0$). However, in the outer region (close to the ansae or handle of the bar), $A_{\Sigma} (x, z)$ still displays non-zero values, thereby indicating that the vertical asymmetry still persists in the outer region. Around $t \sim 6.85 \Gyr$, the model undergoes a second buckling event (see the second peak in $A_{1z}$ albeit with smaller values in Fig.~\ref{fig:temporal_A1z_allmodels}). As a result, at a later time ($t = 6.95 \Gyr$), the model still shows non-zero values for $A_{\Sigma} (x, z)$ in the outer region. By the end of the simulation run ($t = 9 \Gyr$), the values of $A_{\Sigma} (x, z)$ become close to zero throughout the entire region, thereby demonstrating that the vertical symmetry is restored finally. As Fig.~\ref{fig:temporal_verticalasymmetry_anothermethod} clearly reveals that the thin-disc stars show a larger degree of vertical asymmetry (or equivalently larger values of $A_{\Sigma} (x, z)$) when compared with the thick-disc stars.  We further checked the distribution of $A_{\Sigma} (x, z)$ in the $(x, z)$-plane at different times for other thin+thick models which host a prominent b/p structure. We found an overall  similar trend of spatio-temporal evolution of the $A_{\Sigma} (x, z)$ as seen for the model rthickE0.5. 

%##################
% Begin figure
%##################
\begin{figure*}
\centering
\resizebox{\linewidth}{!}{\includegraphics{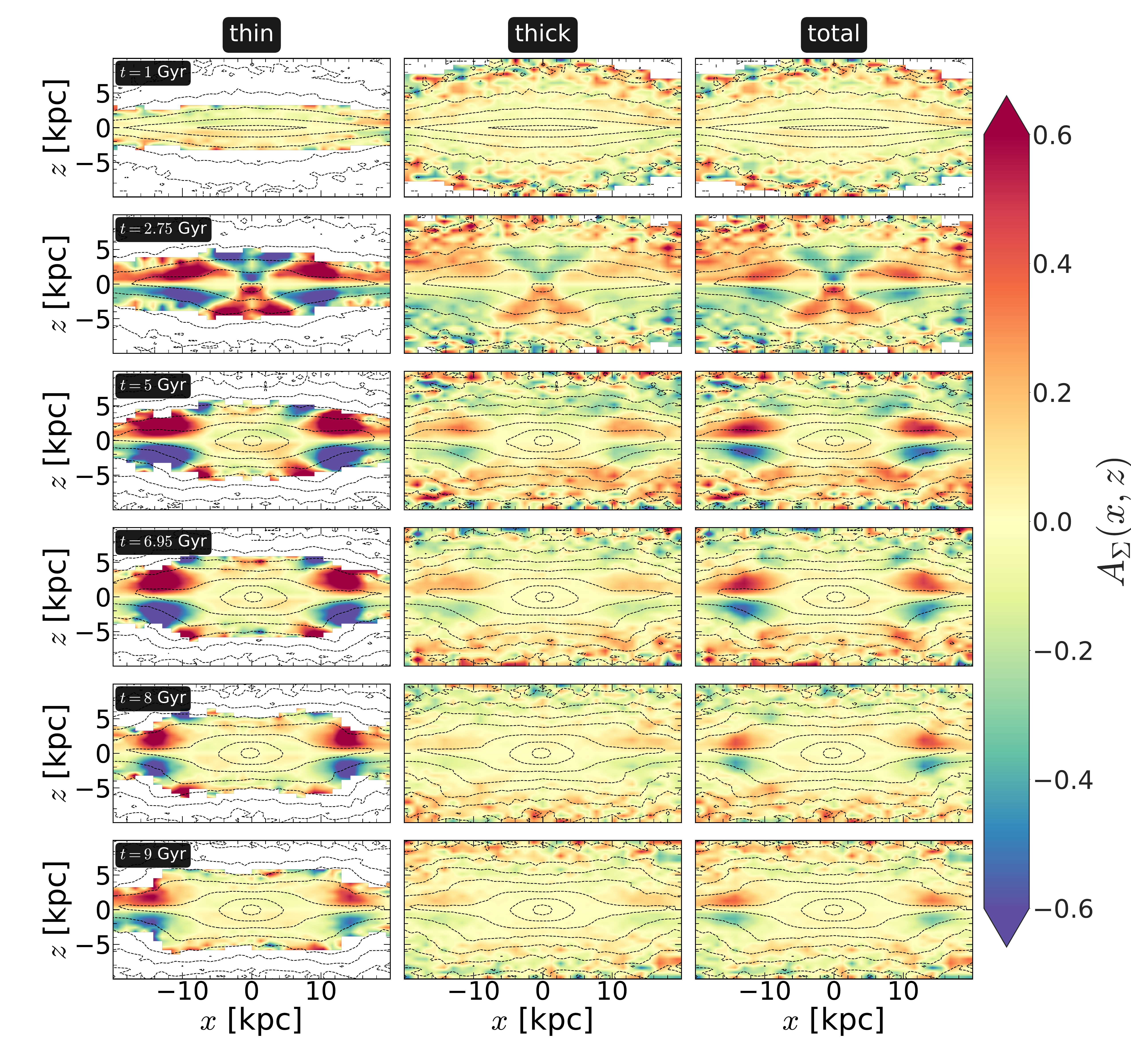}}
\caption{Distribution of the mid-plane asymmetry, $A_{\Sigma} (x, z)$, in the edge-on projection ($x-z$-plane), computed separately for the thin (\textit{left columns}) and thick (\textit{middle columns}) discs, as well as thin+thick (\textit{right columns}) disc particles using Eq.~\ref{eq:midplane_asymmetry} at six different times (before and after the buckling event) for the model rthickE0.5. The bar is placed along the $x$-axis (side-on configuration) for each time-step. Black lines denote contours of constant density. A mid-plane asymmetry persists even long after the model has gone through the bucking phase.}
\label{fig:temporal_verticalasymmetry_anothermethod}
\end{figure*}
%##################
% End figure
%##################

\subsection{Correlation between bar and b/p properties}
\label{sec:bp_bar_correlation}
%&&&&&&&&&&&&&&&&&&&&&&&&&&&&&

Past theoretical studies of the b/p formation and its subsequent growth have revealed a strong correlation between the (maximum) bar strength and the resulting (maximum) b/p strength \citep[e.g. see][]{Martinez-Valpuesta2008}. Here, we test this correlation for the suite of thin+thick models considered here. The maximum bar strengths for the models are obtained from \citet{Ghoshetal2022} where we studied, in detail, the bar properties for these models. The maximum b/p strengths for the models are obtained from Eq.~\ref{eq:bp_strength}. We mention that all stellar particles (thin+thick) are used in calculating the maximum bar and b/p strengths for all models. The resulting distribution of the thin+thick models in maximum bar-maximum b/p strengths are shown in Fig.~\ref{fig:bp_bar_strength_correc}. As seen clearly from Fig.~\ref{fig:bp_bar_strength_correc}, indeed a stronger bar in a model produces a stronger b/p structure. This correlation holds true for all three geometric configurations considered here. Therefore, we find a strong correlation as well between the (maximum) bar strength and the resulting (maximum) b/p strength in our thin+thick models, in agreement with past findings. In addition, we investigated the correlation (if any) between the lengths of the bar and the b/p in our thin+thick models. In Fig.~\ref{fig:bp_bar_length_correc} (top panel), we show the temporal evolution of the ratio of the b/p length ($R_{\rm b/p}$) and the bar length ($R_{\rm bar}$) for the model rthickE0.5. The bar length, $R_{\rm bar}$ is defined as the radial location where the amplitude of the $m=2$ Fourier moment ($A_2/A_0$)  drops to 70 percent of its peak value \citep[for a detailed discussion, see recent work by][]{GhoshandDiMatteo2023}. As seen clearly, the ratio increases shortly after the formation of b/p, and the ratio almost saturates by the end of the simulation run ($9 \Gyr$). Furthermore, we calculated the the b/p length ($R_{\rm b/p}$) and the bar length ($R_{\rm bar}$), at the end of the simulation ($9 \Gyr$) for all thin+thick models considered here. This is shown in Fig.~\ref{fig:bp_bar_length_correc} (bottom panel). For the rthickE and rthickG models, the ratio increases progressively with increasing $f_{\rm thick}$ values. However, for the rthickS model, the ratio increases monotonically till $f_{\rm thick} =0.7$, and then starts to decrease.
\par
Lastly, we investigate the time delay between the bar formation and the onset of buckling instability for all thin+thin models considered here, and study if and how it varies with thick-disc mass fraction ($f_{\rm thick}$). In Appendix~\ref{appen:bar_formation_delay}, we show how the bar formation epoch, $\tau_{\rm bar}$, varies with different $f_{\rm thick}$ values and with different disc configurations.
Similarly, in Sect.~\ref{sec:vert_asymmetry}, we defined the epoch of buckling instability when the peak in $A_{\rm buck}$ occurs. The resulting variation of the time delay, $\tau_{\rm buck} - \tau_{\rm bar}$ with $f_{\rm thick}$ is shown in Fig.~\ref{fig:bp_bar_timedelay}. For a fixed geometric configuration (rthickE, rthickS, or rthickG), the time interval between the bar formation and the onset of buckling instability gets progressively shorter with increasing $f_{\rm thick}$ values. This happens due to the fact that, with increasing $f_{\rm thick}$, the bar forms progressively at a later stage (see Appendix~\ref{appen:bar_formation_delay}). In addition, for a fixed $f_{\rm thick}$ value, the rthickS models almost always show shorter time delay ($\tau_{\rm buck} - \tau_{\rm bar}$) when compared to other two geometric configurations considered here (see Fig.~\ref{fig:bp_bar_timedelay}).

%##################
% Begin figure
%##################
\begin{figure}
\centering
\resizebox{1.0\linewidth}{!}{\includegraphics{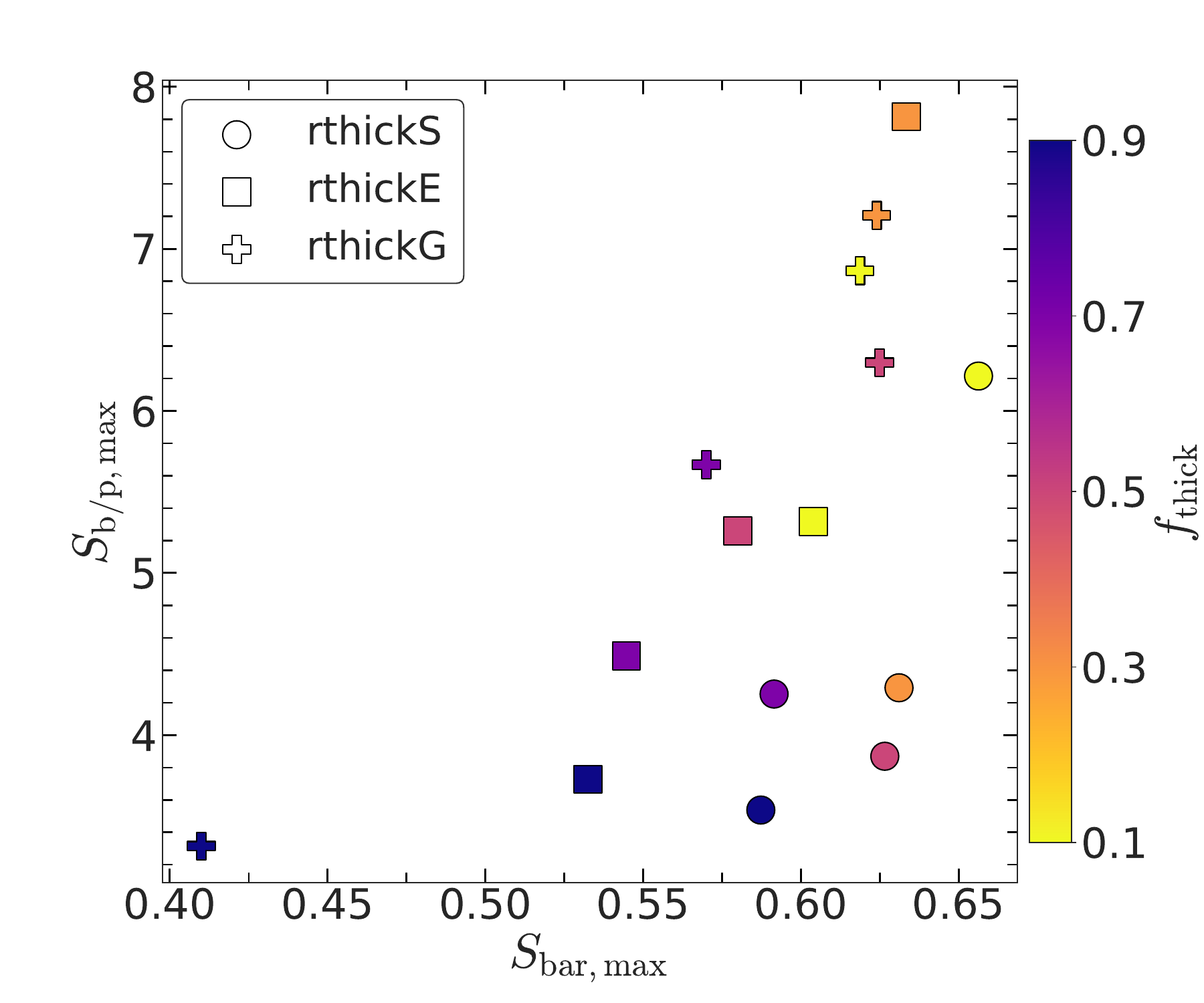}}
\caption{Bar-b/p strength correlation : distribution of all thin+thick models in maximum bar strength -- maximum b/p strength plane. The maximum bar strength are taken from \citet{Ghoshetal2022} whereas the maximum b/p strengths are determined from Eq.~\ref{eq:bp_strength}. The colour bar denotes the thick-disc mass fraction ($f_{\rm thick}$). Different symbols represent models with different geometric configurations (see the legend). The maximum strength of the bar correlates overall with the maximum b/p strength, and this remains true for all three geometric configurations. }
\label{fig:bp_bar_strength_correc}
\end{figure}
%##################
% End figure
%##################

%##################
% Begin figure
%##################
\begin{figure}
\centering
\resizebox{0.95\linewidth}{!}{\includegraphics{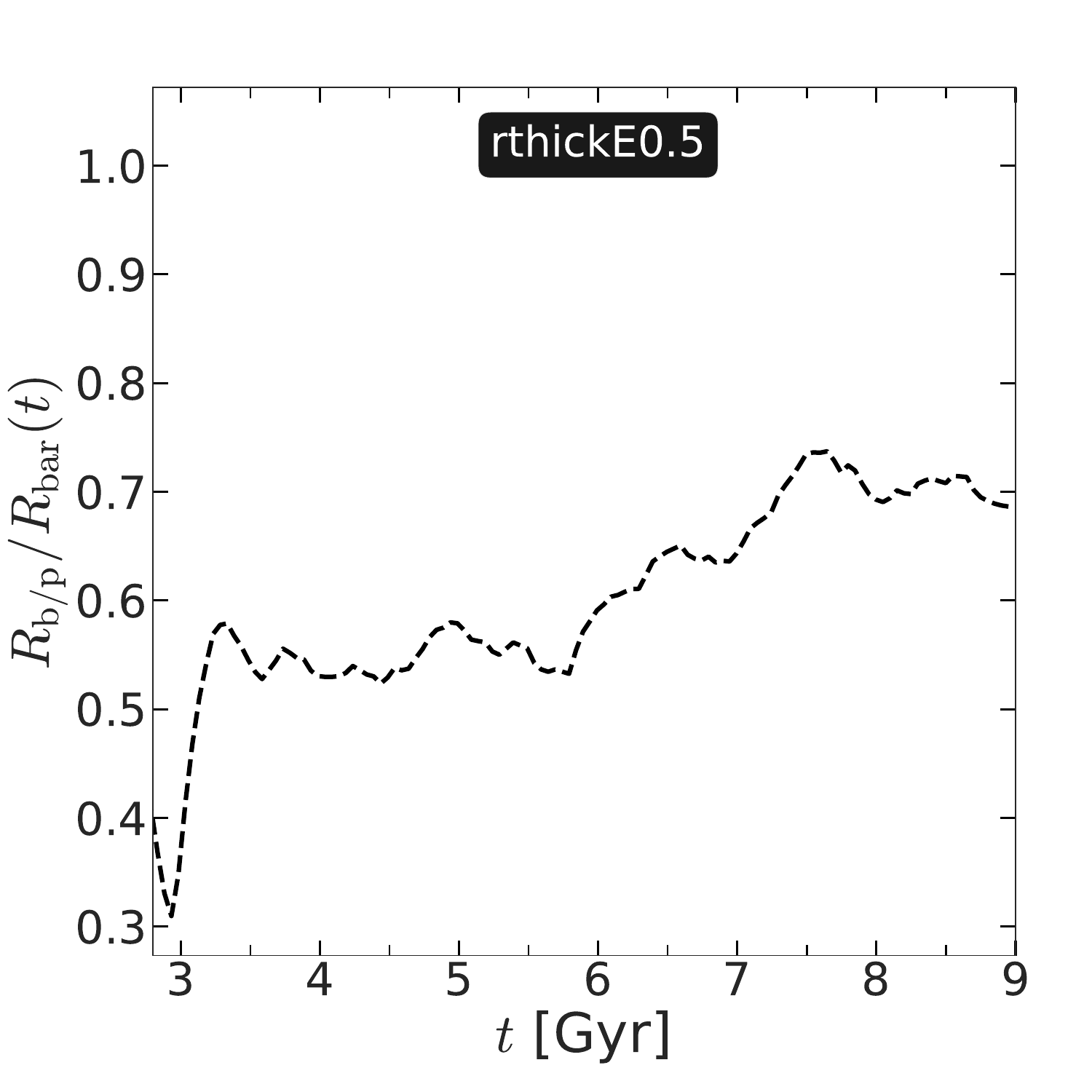}}
\medskip
\resizebox{0.95\linewidth}{!}{\includegraphics{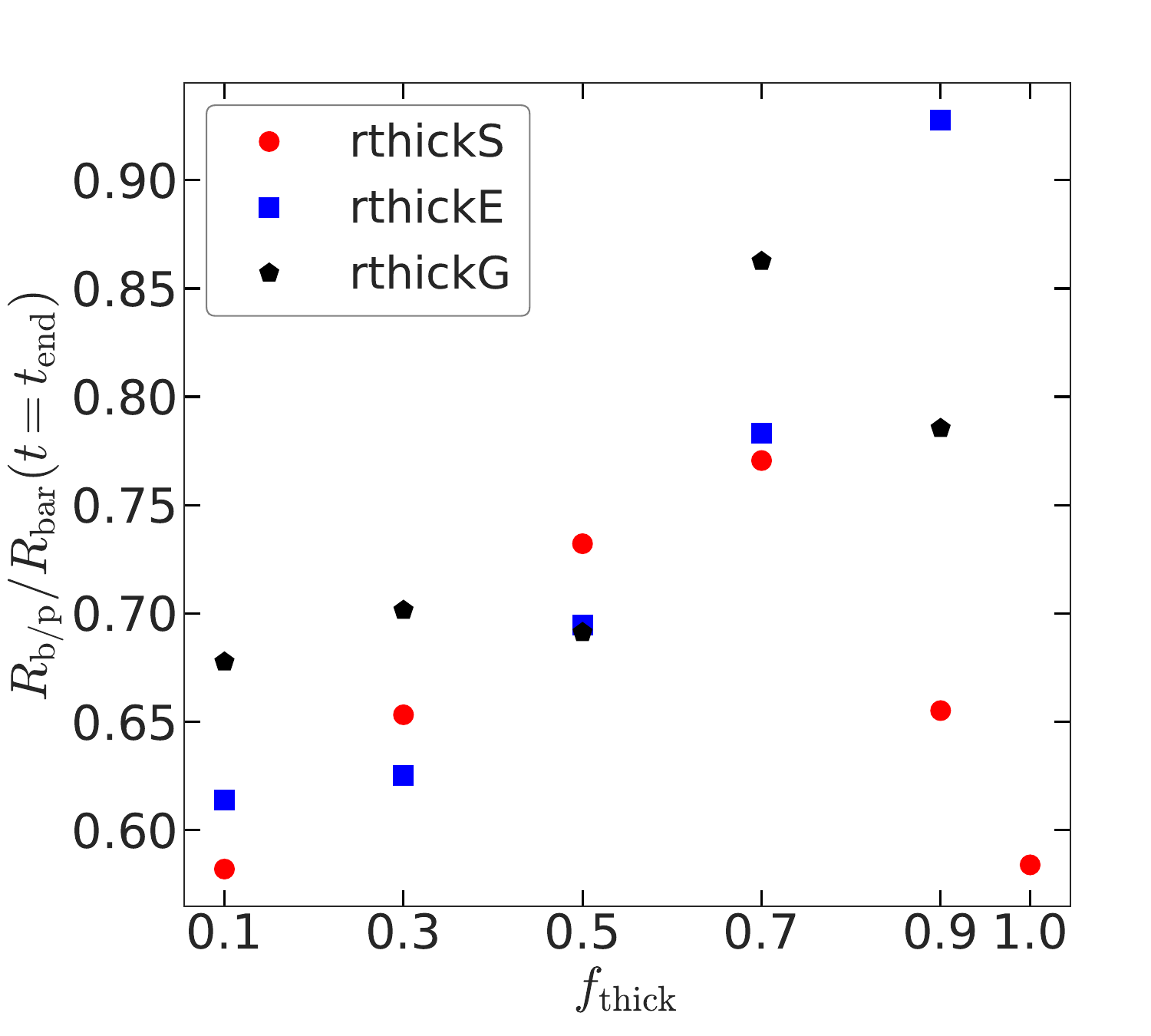}}
\caption{\textit{Top panel}: temporal evolution of the ratio of the b/p length ($R_{\rm b/p}$) and the bar length ($R_{\rm bar}$) for the model rthickE0.5. \textit{Bottom panel}: variation of the ratio of the b/p and the bar length, calculated at the end of the simulation run ($t= 9 \Gyr$), with thick-disc mass fraction ($f_{\rm thick}$). }
\label{fig:bp_bar_length_correc}
\end{figure}
%##################
% End figure
%##################

%##################
% Begin figure
%##################
\begin{figure}
\centering
\resizebox{0.95\linewidth}{!}{\includegraphics{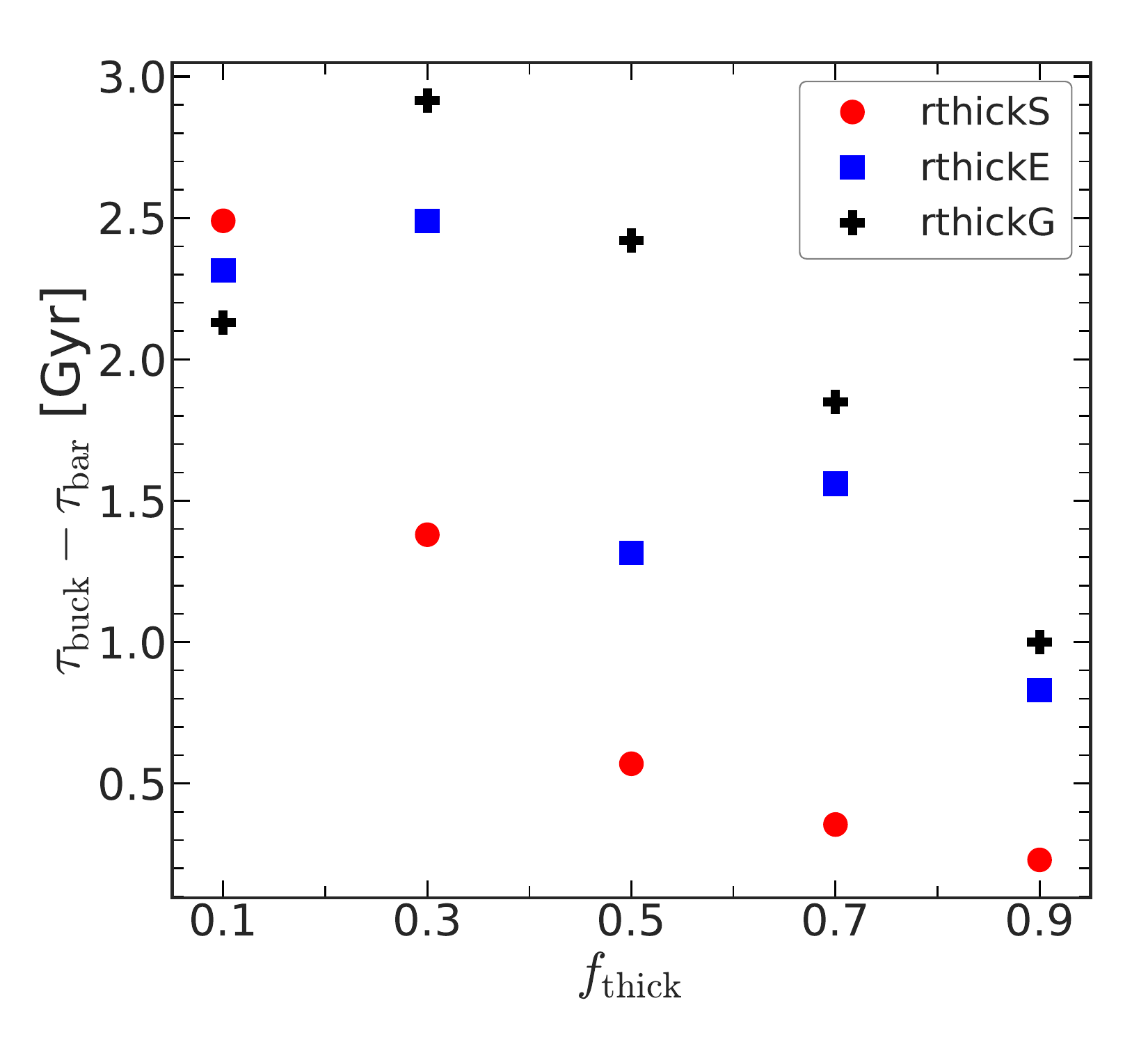}}
\caption{Variation of the time delay between the bar formation and the onset of buckling instability, $\tau_{\rm buck} - \tau_{\rm bar}$, with thick-disc mass fraction ($f_{\rm thick}$), for all thin+thick models considered here.  For a fixed geometric configuration, $\tau_{\rm buck} - \tau_{\rm bar}$ gets progressively shorter with increasing $f_{\rm thick}$ values. For details, see the text. }
\label{fig:bp_bar_timedelay}
\end{figure}
%##################
% End figure
%##################

\section{Kinematic signatures of buckling and its connection with b/p formation}
\label{sec:kinematics}
%&&&&&&&&&&&&&&&&&&&&&&&&&&&&&&&&&

%
%##################
% Begin figure
%##################
\begin{figure*}
\centering
\resizebox{\linewidth}{!}{\includegraphics{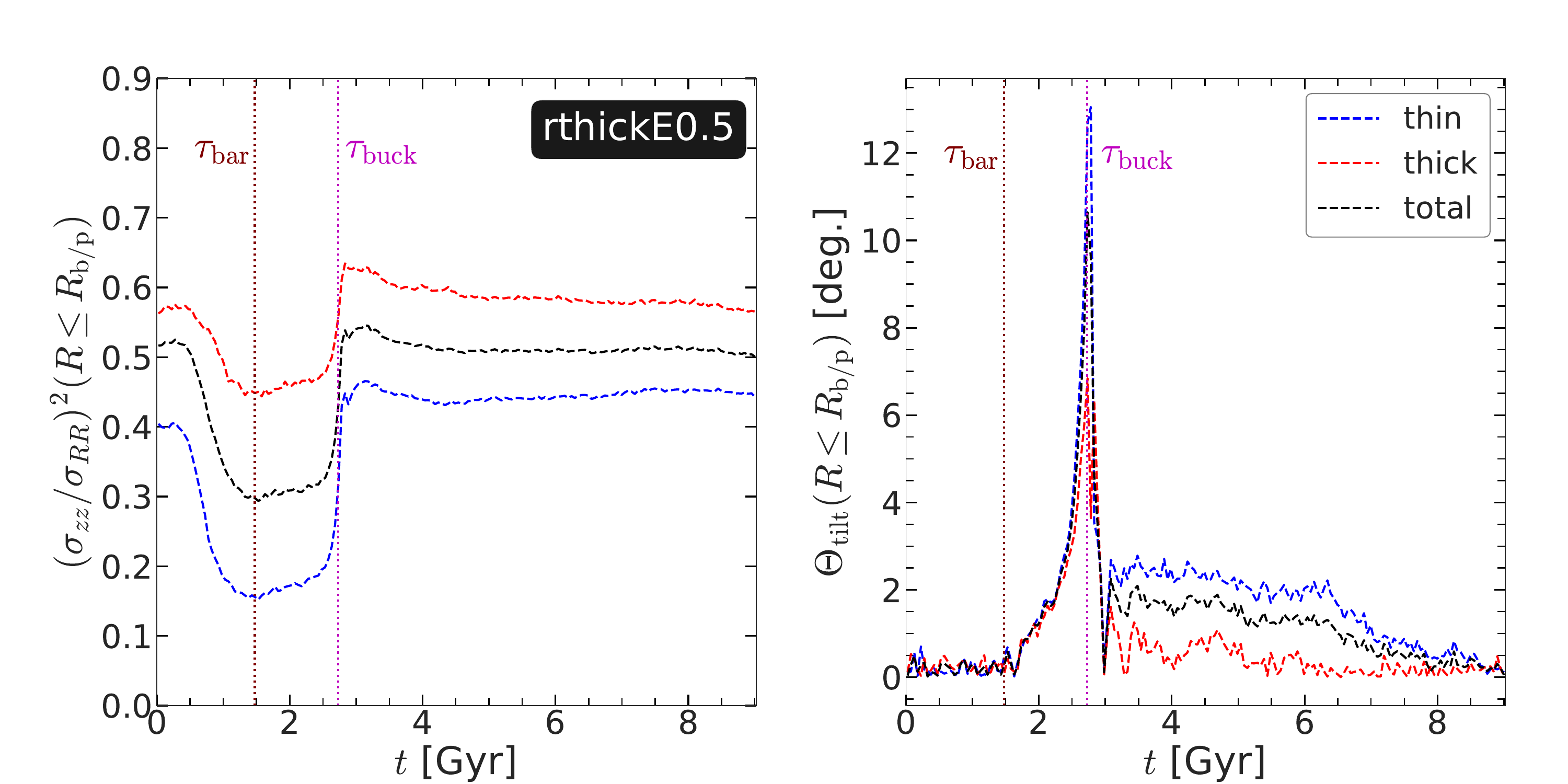}}
\caption{\textit{Left panel :} temporal evolution of the vertical-to-radial velocity dispersion ($\sigma_{zz}/\sigma_{RR}$), calculated within the b/p extent ($R_{\rm b/p}$), $(\sigma_{zz}/\sigma_{RR})^2 (t; R \leq R_{\rm b/p})$, for the thin (in blue), thick (in red), and total (thin+thick) disc (in black) particles, for the model rthickE0.5. \textit{Right panel :} temporal evolution of the meridional tilt angle ($\Theta_{\rm tilt}$), calculated within $R_{\rm b/p}$ using Eq.~\ref{eq:tilt_angle}, for the thin (in blue), thick (in red), and total (thin+thick) disc (in black) particles, for the model rthickE0.5. The vertical maroon dotted line denotes the onset of bar formation ($\tau_{\rm bar}$) while the vertical magenta dotted line denotes the onset of buckling instability ($\tau_{\rm buck}$). For details, see the text.}
\label{fig:rsigmazoversigmaR_signlemodel}
\end{figure*}
%
%##################
% End figure
%##################

Understanding the temporal evolution of the diagonal and off-diagonal components of the stellar velocity tensor was shown to be instrumental for investigating the formation and growth of the b/p structure \citep[e.g., see][and references therein]{Sahaetal2013}. Here, we systematically study some key diagonal and off-diagonal components of the stellar velocity dispersion tensor and their associated temporal evolution for all the thin+thick models considered here.
\par
At a given radius $R$, the stellar velocity dispersion tensor is defined as \citep{BinneyTremaine2008}
\begin{equation}
\sigma^2_{ij} = \avg{v_i v_j} - \avg{v_i} \avg{v_j}\,,
\label{eq:veldisp}
\end{equation}
\noindent where quantities within the $\left < \right>$ bracket denote the average quantities, and the averaging is done for a group of stars.  Here, $i, j = R, \phi, z$. The corresponding stress tensor of the stellar fluid
 is defined as 
\begin{equation}
\tau = \tau_n + \tau_s = - \rho(R) \sigma^2\,,
\label{eq:stress_tensor}
\end{equation}
\noindent where $\rho(R)$ denotes the local volume density of stars at a radial location $R$. $\tau_n$ and $\tau_s$ denote the normal stress (acting along the normal to a small differential imaginary surface $dS$), and the shear stress (acting along the perpendicular to the normal to $dS$), respectively \citep[for further details, see e.g.][]{BinneyTremaine2008,Sahaetal2013}. The components of $\tau_n$ are determined by the diagonal elements of the velocity dispersion tensor while the shear stress are determined by the off-diagonal elements of the velocity dispersion tensor \citep[for details, see][]{BinneyTremaine2008}. Furthermore, the diagonal elements of the velocity dispersion tensor determines the axial ratios of the stellar velocity ellipsoid with respect to the galactocentric axes $(\hat{e}_{R}, \hat{e}_{\phi}, \hat{e}_{z})$ whereas the orientations of the velocity ellipsoid are determined by the off-diagonal elements of the velocity dispersion tensor  \citep[for details, see][]{BinneyTremaine2008,Sahaetal2013}. One such quantity of interest is the meridional tilt angle which is defined as 
\begin{equation}
\Theta_{\rm tilt} = \frac{1}{2} \tan^{-1} \left( \frac{2 \sigma^2_{Rz}}{\sigma^2_{RR} - \sigma^2_{zz}}\right)\,.
\label{eq:tilt_angle}
\end{equation}
\par
The tilt angle, $\Theta_{\rm tilt}$ denotes the orientation or the deformation of the stellar velocity ellipsoid in the  meridional plane ($R-z$-plane).
In the past, it was shown for an $N$-body model that when the bar grows, it causes much radial heating (or equivalently, increasing the radial velocity dispersion, $\sigma_{RR}$) without causing a similar degree of heating the vertical direction (or equivalently, no appreciable increase in the vertical velocity dispersion, $\sigma_{zz}$). Consequently, the model goes through a vertical buckling instability causing the thickening of the inner part, which in turn, also increases $\sigma_{zz}$ \citep[e.g., see][]{Debattistaetal2004,Martinez-Valpuestaetal2006,Martinez-Valpuesta2008,Sahaetal2013,Fragkoudietal2017,DiMatteoetal2019}. Therefore, it is of great interest to investigate the temporal evolution of the vertical-to-radial velocity dispersion ($\sigma_{zz} / \sigma_{RR}$) in order to fully grasp the formation and growth of b/p structure in our thin+thick models. In addition, using $N$-body simulations, \citet{Sahaetal2013} demonstrated that during the onset of the buckling phase, the model shows characteristic increase in the meridional tilt angle, $\Theta_{\rm tilt}$ which in turn, could be used as an excellent diagnostic to identify ongoing buckling phase in real observed galaxies. Here, we study the temporal evolution of these dynamical quantities in detail for all the thin+thick models considered here.
\par
In Appendix~\ref{appen:Veldisp_example}, we show the radial profiles of vertical-to-radial velocity dispersion, as a function of time, for the model rthickE0.5. In order to quantify the temporal evolution of $\sigma_{zz} / \sigma_{RR}$, we compute them using Eq.~\ref{eq:veldisp} within the extent of the b/p structure ($R_{\rm b/p}$). The corresponding temporal evolution of the vertical-to-radial velocity dispersion ($\sigma_{zz} / \sigma_{RR}$), calculated separately for thin, thick, and thin+thick particles, is shown in Fig.~\ref{fig:rsigmazoversigmaR_signlemodel} (left panel) for the model rthickE0.5. As seen from Fig.~\ref{fig:rsigmazoversigmaR_signlemodel}, the temporal evolution of $\sigma_{zz} / \sigma_{RR}$ displays a characteristic `U'-shape (of different amplitudes) during the course of the evolution, arising from the radial heating of the bar (increase in $\sigma_{RR}$) and the subsequent vertical thickening (increase in $\sigma_{zz}$) due to the buckling instability.  In addition, the temporal profiles of $\sigma_{zz} / \sigma_{RR}$ for the thin-disc shows a larger and more prominent `U'-shaped feature when compared with that for the thick-disc stars. This is consistent with the picture that thin disc b/p are, in general, stronger than the thick disc b/p. The epoch corresponding to the maximum increase in the quantity $\sigma_{zz} / \sigma_{RR}$ coincides with the peak in the $A_{\rm buck}$, denoting strong vertical buckling instability (see the location of vertical magenta line in Fig.~\ref{fig:rsigmazoversigmaR_signlemodel}). This further enunciates that the b/p structures in our thin+thick models are indeed formed through vertical buckling instability. In Appendix.~\ref{appen:Veldisp_example}, we show the  temporal evolution of $\sigma_{zz} / \sigma_{RR}$ for all thin+thick models considered here. We checked the the trends, mentioned above, hold  true for all the models which formed a b/p structure during the evolutionary trajectory.
\par
Lastly, we investigate the temporal evolution of the meridional tilt angle, $\Theta_{\rm tilt}$ for the model rthickE0.5. Fig.~\ref{fig:rsigmazoversigmaR_signlemodel} (right panel) shows the corresponding temporal evolution of the tilt angle, $\Theta_{\rm tilt}$, calculated separately for thin, thick, and thin+thick particles using Eq.~\ref{eq:tilt_angle}, for  the model rthickE0.5. The temporal evolution of $\Theta_{\rm tilt}$ shows a characteristic increase during the course of evolution. The epoch of maximum value of the tilt angle coincides with the epoch of strong buckling instability (see the location of vertical magenta line in Fig.~\ref{fig:rsigmazoversigmaR_signlemodel}). This is in agreement with the findings of \citet{Sahaetal2013}, and is consistent with a `b/p formed through buckling' scenario. Furthermore, the temporal profiles of $\Theta_{\rm tilt}$ for the thin disc shows a larger and more prominent peak when compared with that for the thick disc stars. This is expected as the thin disc b/p is stronger than the thick disc b/p. We checked that the trends, mentioned above, hold  true for all the models which formed a b/p structure during the evolutionary trajectory. For the sake of brevity, they are not shown here.
\par
%

%&&&&&&&&&&&&&&&&&&&&&&

\section{X-shape of the b/p and relative contribution of thin disc}
\label{sec:Xshape_example}
%&&&&&&&&&&&&&&&&&&&&&&&&&&&&&&&&&&&&&&&&&&&&&

%##################
% Begin figure
%##################
\begin{figure*}
\centering
\resizebox{\linewidth}{!}{\includegraphics{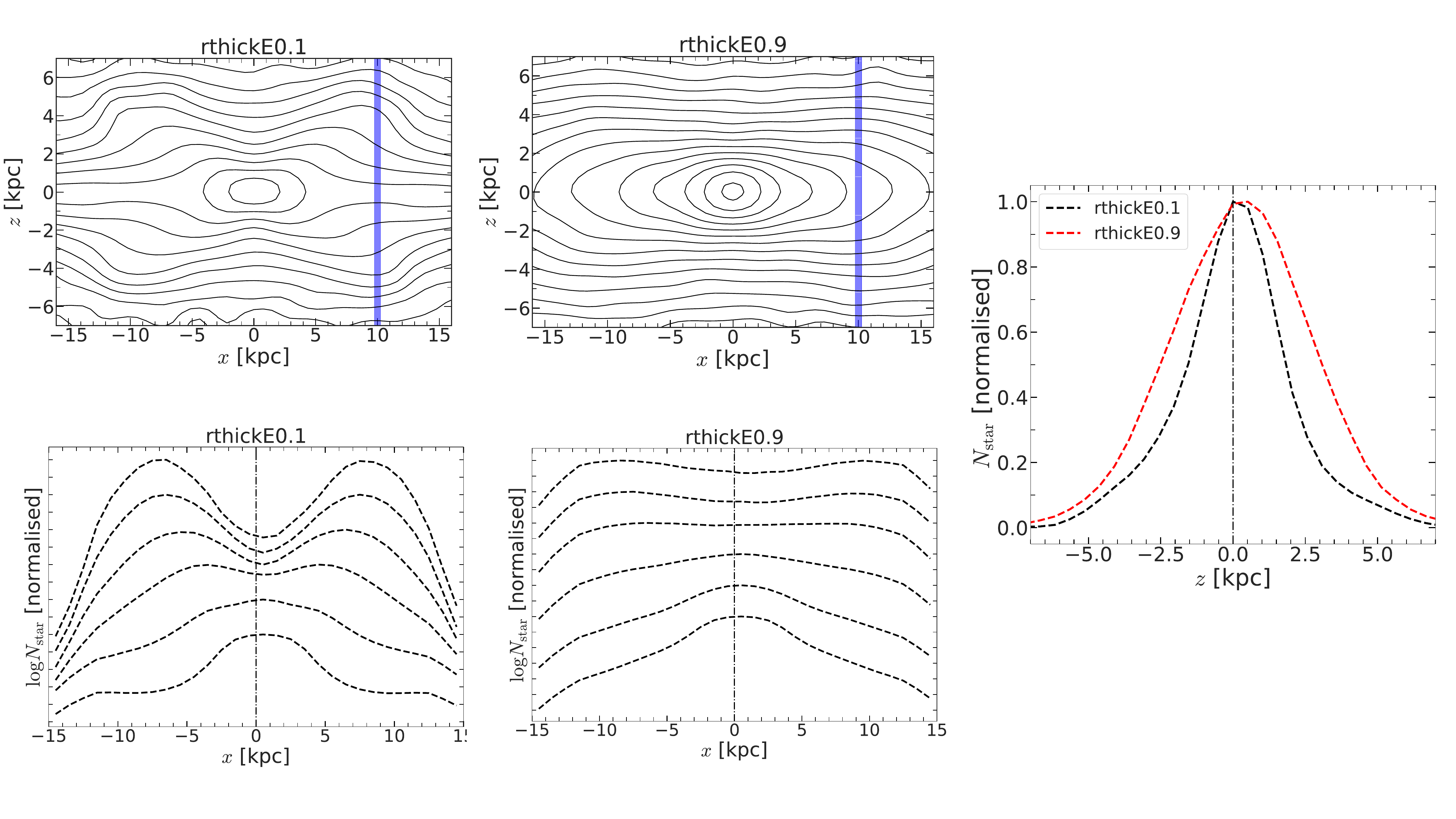}}
\caption{\textit{Top left panels :} density contours of the edge-on stellar (thin+thick) distribution (with bar placed along the $x$-axis) in the central region at $t = 9 \Gyr$ for the models rthickE0.1 and rthickE0.9. For the rthickE0.1 model, the contours display more prominent `X-shaped' feature whereas for the rthickE0.9 model, the contours display more prominent `boxy-shaped' feature. \textit{Bottom left panels :} Density profiles (normalised by the peak density value, and in log-scale) along the bar major axis, calculated at different heights (from $|z| = 0 $ to $6 \kpc$, with a step-size of $1 \kpc$) from the mid-plane, for the models rthickE0.1 and rthickE0.9. The density profiles have been artificially shifted along the $y$-axis to show the trends as height changes and not overlap. \textit{Right panel :} Vertical stellar density distribution (normalised by the peak density value, and in log-scale), at a radial location around the peak of the b/p structure (marked by blue vertical lines in \textit{top left panels}) for the rthickE0.1 and rthickE0.9 models.}
\label{fig:Xshape_to_boxy_demo}
\end{figure*}
%##################
% End figure
%##################

%##################
% Begin figure
%##################
\begin{figure*}
\centering
\resizebox{\linewidth}{!}{\includegraphics{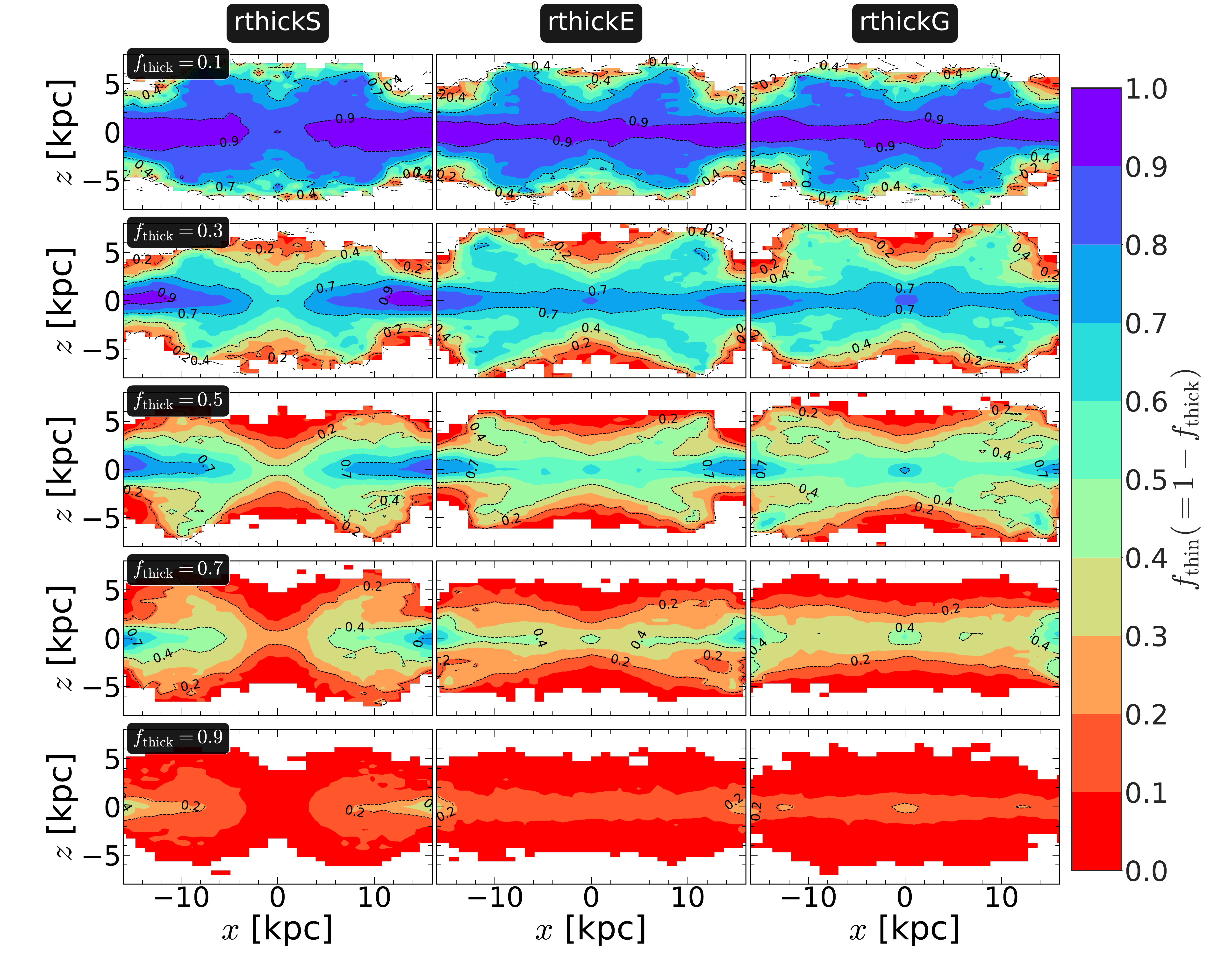}}
\caption{Fraction of thin-disc stars, $f_{\rm thin} (= 1 - f_{\rm thick})$, in the edge-on projection (with the bar placed along the $x$-axis) compared to the total (thin+thick) disc, at the end of the simulation run ($t = 9 \Gyr$) for all thin+thick disc models with varying $f_{\rm thick}$ values. \textit{Left panels} show the distribution for the rthickS models whereas  \textit{middle panels} and \textit{right panels} show the distribution for the rthickE  and rthickG models, respectively. The values of ($f_{\rm thick}$) varies from 0.1 to 0.9 (top to bottom), as indicated in the left-most panel of each row. For each model, the fraction of thin disc stars decreases with height from the mid-plane ($z =0$). In addition, the appearance of the b/p structure changes from more `X-shaped' to more `boxy-shaped' as the thick-disc mass fraction steadily increases.}
\label{fig:thindiscfraction_maps_endstep_allmodels}
\end{figure*}
%##################
% End figure
%##################

A visual inspection of Fig.~\ref{fig:density_maps_endstep_allmodels} already revealed that for a fixed geometric configuration, the appearance of the b/p structure changes from more `X-shaped' to more `boxy-shaped' as the thick-disc mass fraction steadily increases. This trend is more prominent for the rthickE and rthickG models (see middle and right panels of Fig.~\ref{fig:density_maps_endstep_allmodels}). Here, we investigate this in further details. In addition, we also investigate how the thin- and thick-disc stars contribute to the formation of the b/p structure.
\par
To carry out the detailed analysis, we first choose two models, namely, rthickE0.1 and rthickE0.9. In the rthickE0.1 model, the thin-disc stars dominate the underlying mass distribution whereas in the rthickE0.9 model, the thick-disc stars dominate the mass distribution, thereby providing an ideal scenario for the aforementioned investigation. In Fig.~\ref{fig:Xshape_to_boxy_demo} (\textit{top left panels}), we
 show the density contours of the edge-on stellar (thin+thick) distribution (with bar placed along the $x$-axis) in the central region encompassing the b/p structure. This clearly brings out the stark differences in the morphology of density contours -- in the rthickE0.1 model, the contours have more prominent `X-shaped' appearance whereas in the rthickE0.9 model, the contours have more prominent `boxy-shaped' appearance. Fig.~\ref{fig:Xshape_to_boxy_demo} (\textit{bottom left panels}) shows the density profiles along the bar major axis, calculated at different heights (from the mid-plane) for these two thin+thick models. At larger heights, a  bimodality in the density profiles along the bar major axis reconfirms the strong X-shaped feature for the rthickE0.1 model. For the rthickE0.9 model, no such bimodality in the density profiles along the bar major axis is seen, thereby confirming that the b/p structure is more boxy-shaped in rthickE0.9 model. Furthermore, we calculated the vertical stellar density distribution, at a radial location around the peak of the b/p structure (see vertical blue lines in top left panels of Fig.~\ref{fig:Xshape_to_boxy_demo}) for these two models. A careful inspection reveals that the vertical stellar density distribution for the rthickE0.1 model is more centrally-peaked (with well-defined tails) whereas vertical stellar density distribution for the rthickE0.9 model is broader, especially at larger heights (see right panel of Fig.~\ref{fig:Xshape_to_boxy_demo}).
\par
To further investigate how the thin- and thick-disc stars contribute to the formation of the b/p structure, we calculate the thin-disc mass fraction, $f_{\rm thin} (=1 - f_{\rm thick}$), at the end of the simulation run ($t = 9 \Gyr$). Fig.~\ref{fig:thindiscfraction_maps_endstep_allmodels} shows the corresponding distribution of the thin-disc mass fraction in the edge-on projection ($x-z$-plane) for all thin+thick disc models. In each case, the bar is placed in side-on configuration (along the $x$-axis). As seen clearly from Fig.~\ref{fig:thindiscfraction_maps_endstep_allmodels}, the thin-disc stars dominate in central regions close to the mid-plane ($z =0$), and are responsible for giving rise to a strong X-shape of the b/p structure, in agreement with the findings of past studies \citep[see e.g. ][]{DiMatteo2016,Athanassoulaetal2017,Debattistaetal2017,Fragkoudietal2017,Fragkoudietal2020}. As one moves farther away from the mid-plane, the thick-disc stars start to dominate progressively. Furthermore, the appearance of the b/p structure changes from more `X-shaped' to more `boxy-shaped' as the thick-disc mass fraction steadily increases. These trends remain generic for all three geometric configurations (different thin-to-thick disc scale length ratios) considered here.

\section{Discussion}
\label{sec:discussion}
%&&&&&&&&&&&&&&&&&&&&&&&&&&
In what follows, we discuss some of the implications and limitations of this work.
 First, our findings demonstrate that the b/p structure can form even in the presence of a massive thick-disc component. This provides a natural explanation to the presence of b/p in high redshift ($z =1$) disc galaxies in the hypothesis that these high-z discs have a significant fraction of their mass in a thick disc \citep[see e.g.][]{Hamilton-Camposetal2023}. A recent work by \citet{Kruketal2019} estimated that at $z \sim 1$, about 10 percent of barred galaxies would harbour a b/p. Therefore, our results are in agreement with the recent observational trends. In addition, bars forming in presence of a massive thick-disc \citep[as shown in a recent work of][]{Ghoshetal2022} and the present work showing b/ps form as well  in presence of a massive thick-disc suggest that bars and b/p bulges formation may have appeared at earlier redshifts than what has been considered so far, also in galaxies dominated by a thick disc component. The findings that the b/p morphology and length depend on the thick-disc fraction (the higher the thick-to-thin disc mass ratio, the more boxy the corresponding b/p and the smaller its extent) may be taken as a prediction that can be tested in current and future observations (JWST, for example). 
\par
 Secondly, the occurrence of b/p in disc galaxies is observationally found to be strongly dependent on the stellar mass of the galaxy in the local Universe \citep{YoshinoandYamauchi2015,ErwinandDebattista2017,Marchuketal2022} as well as at higher redshifts \citep{Kruketal2019}. This implies that galaxy's stellar mass is likely to play an important role for forming b/ps via vertical instabilities. In our suite of thin+thick models, although we systematically varied the thick-disc mass fraction, the total stellar mass remained fixed ($\sim 1 \times 10^{11} M_{\odot}$). Investigating the role of the stellar mass on b/p formation via vertical buckling instability would be quite interesting, however it is beyond the scope of the present work.
 \par
 Furthermore, in our thin+thick models, the stars are separated into two well-defined and distinct populations, namely, thin- and thick-disc stars. While this might be well-suited for external galaxies, however this scheme is a simplification for the Milky Way. \citet{Bovyetal2012} showed that the disc properties vary continuously with the scale height in the Milky Way. Nevertheless, our adapted scheme of discretising stars with a varying fraction of thick-disc stars provides valuable insight into the trends as it has been shown for the MW, a two component disc can already capture the main trends found in more complex, multi-component discs \citep[e.g. see][]{DiMatteo2016,Debattistaetal2017,Fragkoudietal2017,Fragkoudi2018,Fragkoudietal2018,Fragkoudietal2020}.
 \par
 Finally, if these b/p structures may have formed also at high redshift, two additional questions arise: the role of interstellar gas which is particularly critical in high-$z$ discs which are gas-rich, and the role of mergers and accretions - which high-$z$ galaxies may have experienced at high rates - in maintaining/perturbing/destroying bars and b/p bulges. The role of the interstellar gas in the context of the generation/destruction of disc instabilities, such as bars \citep{Bournaudetal2005} and spiral arms \citep{SellwoodCarlberg1984,GhoshJog2015,GhoshJog2016,GhoshJog2021}, has been investigated in past literature. In addition, the b/p bulges play a key role in evolution of disc galaxies by regulating the bar-driven gas inflow \citep[e.g., see][]{Fragoudietal2015,Fragoudietal2016}. Furthermore, bars can be weakened substantially (or even destroyed in some cases) as a result of minor mergers \citep{Ghoshetal2021}. Therefore, it would be worth investigating the b/p formation and evolution scenario in the presence of the thick disc and the interstellar gas, and how they are likely to be affected by the merger events.

\section{Summary}
\label{sec:conclusion}
%&&&&&&&&&&&&&&&&&&&&&&&&
In summary, we investigated the dynamical role of a geometrically thick-disc on the b/p formation and their subsequent evolution scenario. We made use of a suite of $N$-body models of thin+thick discs and systematically varied mass fraction of the thick disc and the different thin-to-thick disc scale length ratios. Our main findings are listed below.\\

\begin{itemize}

\item{b/ps form in almost all thin+thick disc models with varying thick disc mass fractions and for all three geometric configurations with different thin-to-thick disc scale length ratios. The thick disc b/p always remains weaker than the thin disc b/p, and this remains valid for all three geometric configurations considered here.}

\item{The final b/p strength shows an overall decreasing trend with increasing thick-disc mass fraction ($f_{\rm thick}$). In addition, the b/ps in simulated galaxies with shorter thick-disc scale lengths form at earlier times and show a rapid initial growth phase when compared to other two geometric configurations. Furthermore, we found a strong (positive) correlation between the maximum bar and b/p strengths in our thin+thick models.}

\item{For a fixed geometric  configuration, the time interval between the bar formation and the onset of vertical buckling instability gets progressively shorter with increasing thick-disc mass fraction. In addition, for a fixed thick-disc mass fraction, models with shorter thick-disc scale length display shorter time delay between the bar formation and the onset of buckling event when compared to other two geometric configurations.}

\item{The final b/p length shows an overall increasing trend with increasing thick-disc mass fraction ($f_{\rm thick}$), and this remains valid for all three geometric configurations considered here. In addition, for a fixed $f_{\rm thick}$ value, the models with larger thick-disc scale lengths form a larger b/p structure when compared to other two geometric configurations. Furthermore, the weaker b/ps are more extended structures (i.e. larger $R_{\rm b/p}$).}

\item{The b/p structure changes appearance from being more `X-shaped' to being more `boxy-shaped' as the $f_{\rm thick}$ values increase monotonically. This trend holds true for all three geometric configurations. Furthermore, the thin-disc stars are predominantly responsible for giving rise to a strong X-shape of the b/p structure.}

\item{Our thin+thick models go through a vertical buckling instability phase to form the b/p structure. The thin-disc stars display a higher degree of vertical asymmetry/buckling when compared to the thick-disc stars. Furthermore, the vertical asymmetry persists long after the buckling phase is over -- the vertical symmetry in the inner region is restored relatively quickly while the vertical symmetry in the outer region (close to the ansae or handle of the bar) is restored long after the buckling event is over.}

\item{The thin+thick models demonstrate characteristic signatures in the temporal evolution of different diagonal ($\sigma_{zz}/\sigma_{RR}$ ratio) and off-diagonal (meridional tilt angle, $\Theta_{\rm tilt}$) components of the stellar velocity dispersion tensor as one would have expected if the b/p structure is formed via the vertical buckling instability. These kinematic signatures are more pronounced/prominent when computed using only the thin-disc stars as compared to using only the thick-disc stars.}

\end{itemize}

To conclude, even in presence of a massive (kinematically hot) thick-disc component, the models are to able to harbour prominent b/p structure formed via vertical bucking instability. This clearly implies that b/ps can form in thick discs at high redshifts, and is in agreement with the observational evidences of presence of b/ps at high redshifts \citep{Kruketal2019}. Our results presented here also predict that at higher redshifts, the b/p will have more `boxy-shaped' appearance than more `X-shaped' appearance which remains to be tested from future observations at higher redshifts ($z =1$ and beyond).

\section*{Acknowledgements}
S.G. acknowledges funding from the Alexander von Humboldt Foundation, through Dr. Gregory M. Green's Sofja Kovalevskaja Award. We thank the anonymous referee for useful comments which helped to improve this paper. This work has made use of the computational resources obtained through the DARI grant A0120410154 (P.I. : P. Di Matteo).

\bibliographystyle{aa.bst} % style aa.bst
\bibliography{my_ref.bib} % your references Yourfile.bib

\begin{appendix}

\section{b/p strength from the Fourier decomposition}
\label{appen:bp_strength_fourier}
%&&&&&&&&&&&&&&&&&&&&&&&&&&&&

%##################
% Begin figure
%##################
\begin{figure}
\centering
\resizebox{\linewidth}{!}{\includegraphics{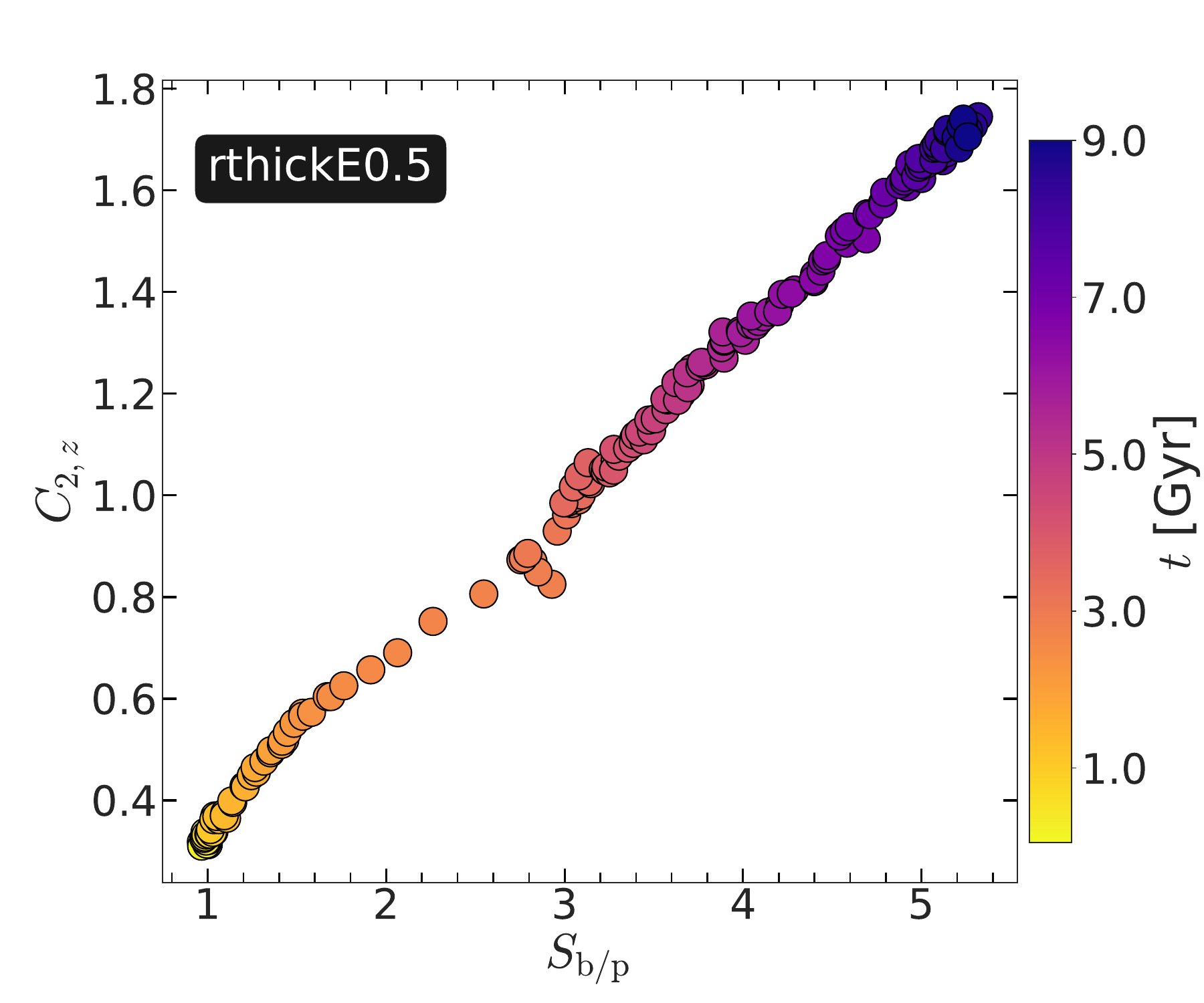}}
\caption{Temporal evolution of  $C_{2, z}$, the b/p strength computed from the Fourier decomposition using Eq.~\ref{eq:bp_strength_fourier}, plotted against the b/p strength calculated using Eq.~\ref{eq:bp_strength} for the model rthickE0.5. All the stellar particles (thin+thick) within the b/p extent ($R_{\rm b/p}$) are used in both cases. The colour bar denotes the time of the simulation. Both the measures of b/p strength correlates fairly well.}
\label{fig:bp_strength_fourier}
\end{figure}
%##################
% End figure
%##################

Following  \citet{Martinez-Valpuestaetal2006} and \citet{Martinez-Valpuesta2008}, at time $t$, we define the strength of the b/p as 

\begin{equation}
C_{m, z} = \left |\sum_{i=1} ^{N_s} z_i \exp[{i m z_i/(5 \avg{z_{\rm d}})}] \right|\,; m =2\,,
\label{eq:bp_strength_fourier}
\end{equation}
\noindent where $z_i$ denotes the vertical position of the $i$th particle, $N_s$ denotes the total number of stellar particles (thin+thick) within the extent of the b/p.  $\avg{z_{\rm d}}$ denotes the average scale height of the total disc, and is calculated in the following way 
\begin{equation}
\avg{z_{\rm d}}  = \frac{M_{\rm d, thin} z_{\rm d, thin} + M_{\rm d, thick} z_{\rm d, thick}} {M_{\rm d, thin}+M_{\rm d, thick}}\,.
\label{eq:avg_scaleheight}
\end{equation}

In Fig.~\ref{fig:bp_strength_fourier}, we show the temporal evolution of $C_{2, z}$, computed using Eq.~\ref{eq:bp_strength_fourier} for the model rthickE0.5 and compare the corresponding b/p strength calculated using Eq.~\ref{eq:bp_strength}. As seen clearly from Fig.~\ref{fig:bp_strength_fourier}, there is an one-to-one correspondence between the b/p strengths, calculated using Eq.~\ref{eq:bp_strength_fourier} and Eq.~\ref{eq:bp_strength}. We checked this for other thin+thick models as well, and found a similar trend as seen for the model rthickE0.5. For the sake of brevity, they are not shown here.

\section{b/p length from the LOS surface density profile}
\label{appen:bp_length_fourier}
%&&&&&&&&&&&&&&&&&&&&&&&&&&&&

\citet{Luttickeetal2000} and \citet{SahaandGerhard2013} measured the size of the b/p structure, $L_{\rm b/p}$, by finding zeros of the
function $D_g (x, z)$ defined by 
\begin{equation}
D_g (x, z) = \frac{\Sigma_{\rm los} (x, z) - \Sigma_{\rm los} (0, z)}{\Sigma_{\rm los} (0, z)}\,,
\label{eq:bp_size_kanak}
\end{equation}
\noindent for a set of smoothed surface density ($\Sigma_{\rm los}$) profiles while placing the bar along the $x$-axis. In Fig.~\ref{fig:bp_length_alternative} (top panel), we show the profiles of the function $D_g (x, z)$ at three different times where the $\Sigma_{\rm los} (x, z)$ profiles are calculated at $z = 0.8 R_{\rm d, thin}$ \citep[for further details, see][]{SahaandGerhard2013}. Fig.~\ref{fig:bp_length_alternative} (bottom panel) shows the correlation of b/p size, $L_{\rm b/p}$ (computed by finding zeros of function $D_g (x, z)$) with the b/p length, $R_{\rm b/p}$. As seen clearly, these two quantities are strongly correlated with the Pearson correlation coefficient, $\rho \sim 0.95$. However, at all times, the values of $L_{\rm b/p}$ remains larger than that of $R_{\rm b/p}$, as also can be judged from the slope of the best-fit straight line (see black dashed line in bottom panel of Fig.~\ref{fig:bp_length_alternative}).

%##################
% Begin figure
%##################
\begin{figure}
\centering
\resizebox{\linewidth}{!}{\includegraphics{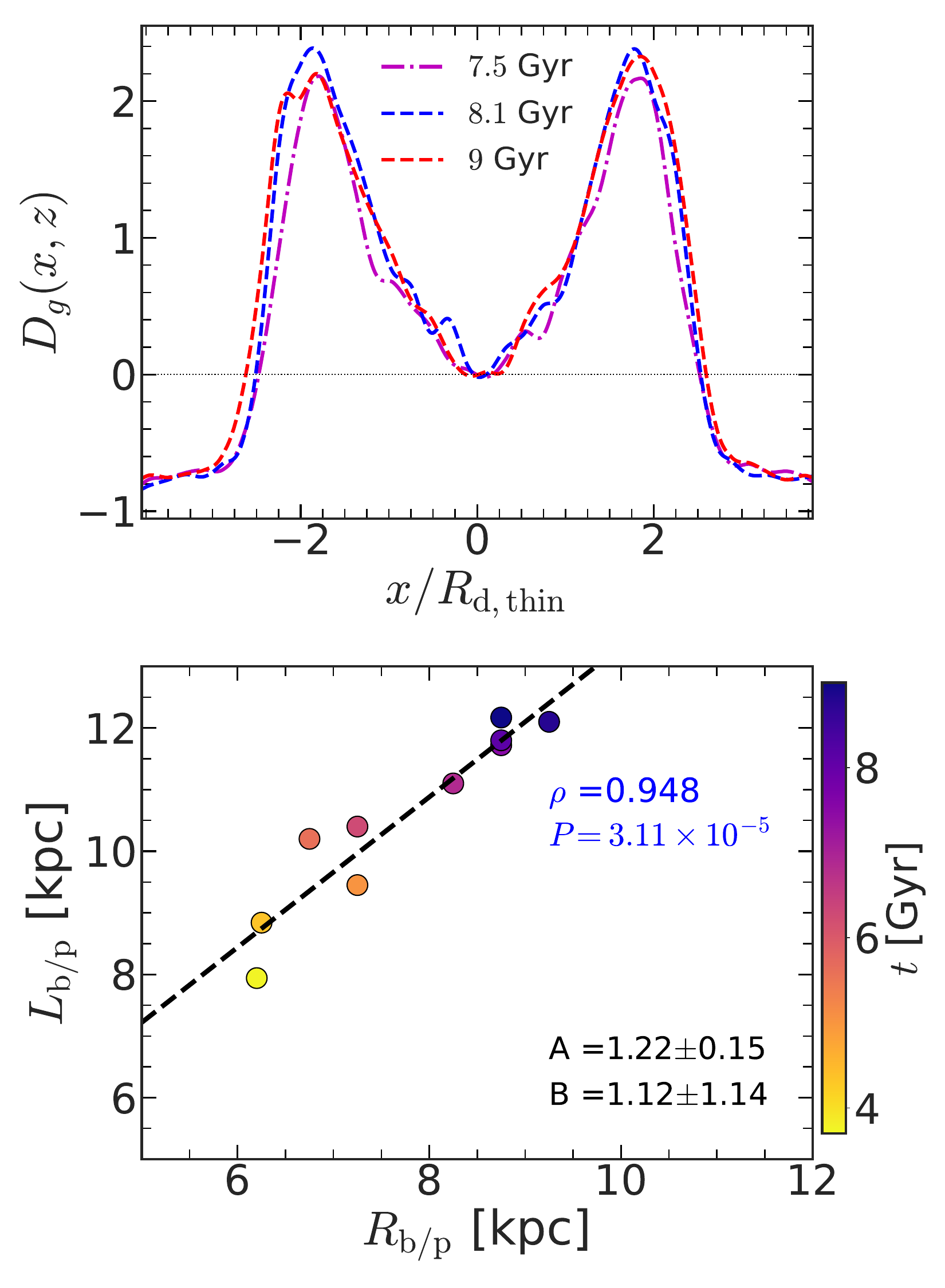}}
\caption{ \textit{Top panel:} profiles of the function $D_g (x, z)$ (see Eq.\ref{eq:bp_size_kanak}), calculated at three different times for the model rthickE0.5. The horizontal dotted line denotes the zeros of the  function $D_g (x, z)$, for details see the text. \textit{Bottom panel :} correlations between the size of the b/p, $L_{\rm b/p}$ and the b/p length $R_{\rm b/p}$ for the model rthickE0.5.  A straight line of the form $Y = AX+B$ is fitted (black dashed line), and the corresponding best-fit parameters are quoted. The Pearson correlation coefficients are quoted (see top right). Here, $R_{\rm d, thin} = 4.7 \kpc$.}
\label{fig:bp_length_alternative}
\end{figure}
%##################
% End figure
%##################

\section{Thin \& thick disc b/p lengths as a function of thick disc mass fraction}
\label{appen:bp_length_componentwise}
%&&&&&&&&&&&&&&&&&&&&&&&&&&&&

%##################
% Begin figure
%##################
\begin{figure*}
\centering
\resizebox{\linewidth}{!}{\includegraphics{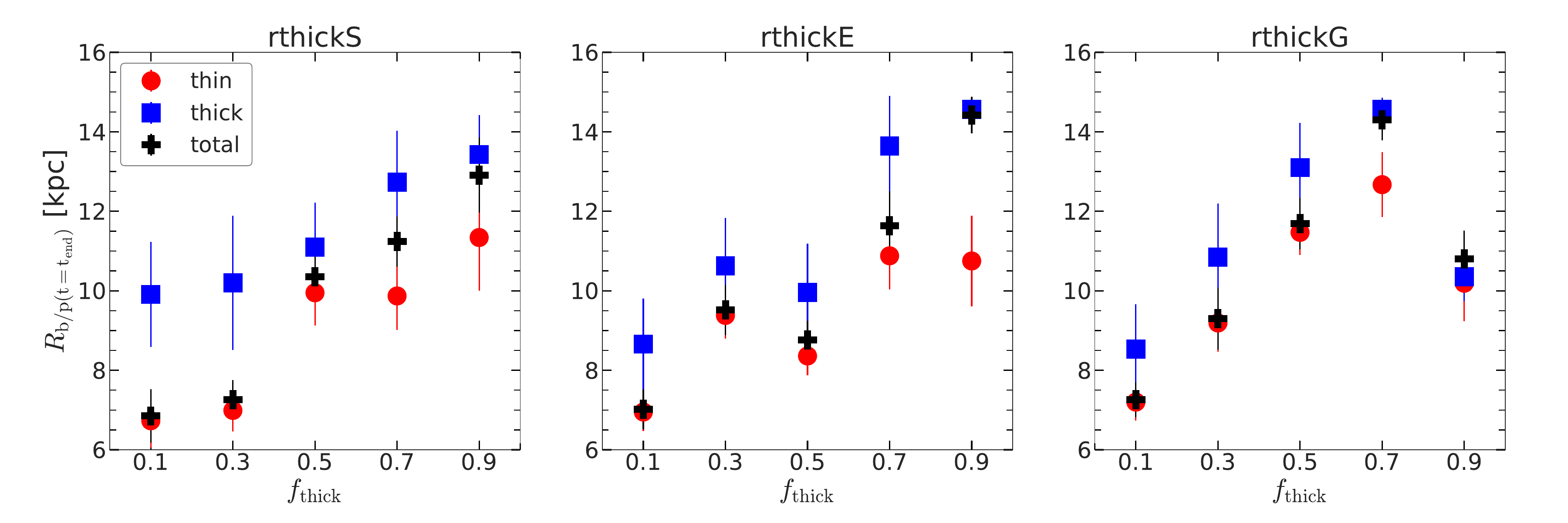}}
\caption{ The b/p extent, at the end of the simulation run ($t = 9 \Gyr$), computed for the thin- and thick-disc, as well as for the thin+thick case, for all the thin+thick models considered here.  \textit{Left panels} show the distribution for the rthickS models whereas  \textit{middle panels} and \textit{right panels} show the distribution for the rthickE  and rthickG models, respectively. Thin disc b/ps always remain shorter than the thick disc b/ps. The errors on $R_{\rm b/p}$ are estimated by constructing a total of 5,000 realisations by resampling the entire population via bootstrapping technique. For details, see Sect.~\ref{sec:bp_length_temporal_evolution}.}
\label{fig:bar_length_componentwise_endstep}
\end{figure*}
%##################
% End figure
%##################

Here, we briefly investigate how the extent of the thin and thick disc b/ps, at the end of the simulation run ($t = 9 \Gyr$) vary with the thick-disc mass fraction in all three geometric configurations considered here. This is shown in Fig.~\ref{fig:bar_length_componentwise_endstep}. For a given thin+thick model, the thin disc b/p always remains shorter than the thick disc b/p, and this holds for almost all thin+thick models considered here.

\section{bar formation epoch and its variation with the thick disc mass fraction}
\label{appen:bar_formation_delay}
%&&&&&&&&&&&&&&&&&&&&&&&&&&&&

%##################
% Begin figure
%##################
\begin{figure}
\centering
\resizebox{0.95\linewidth}{!}{\includegraphics{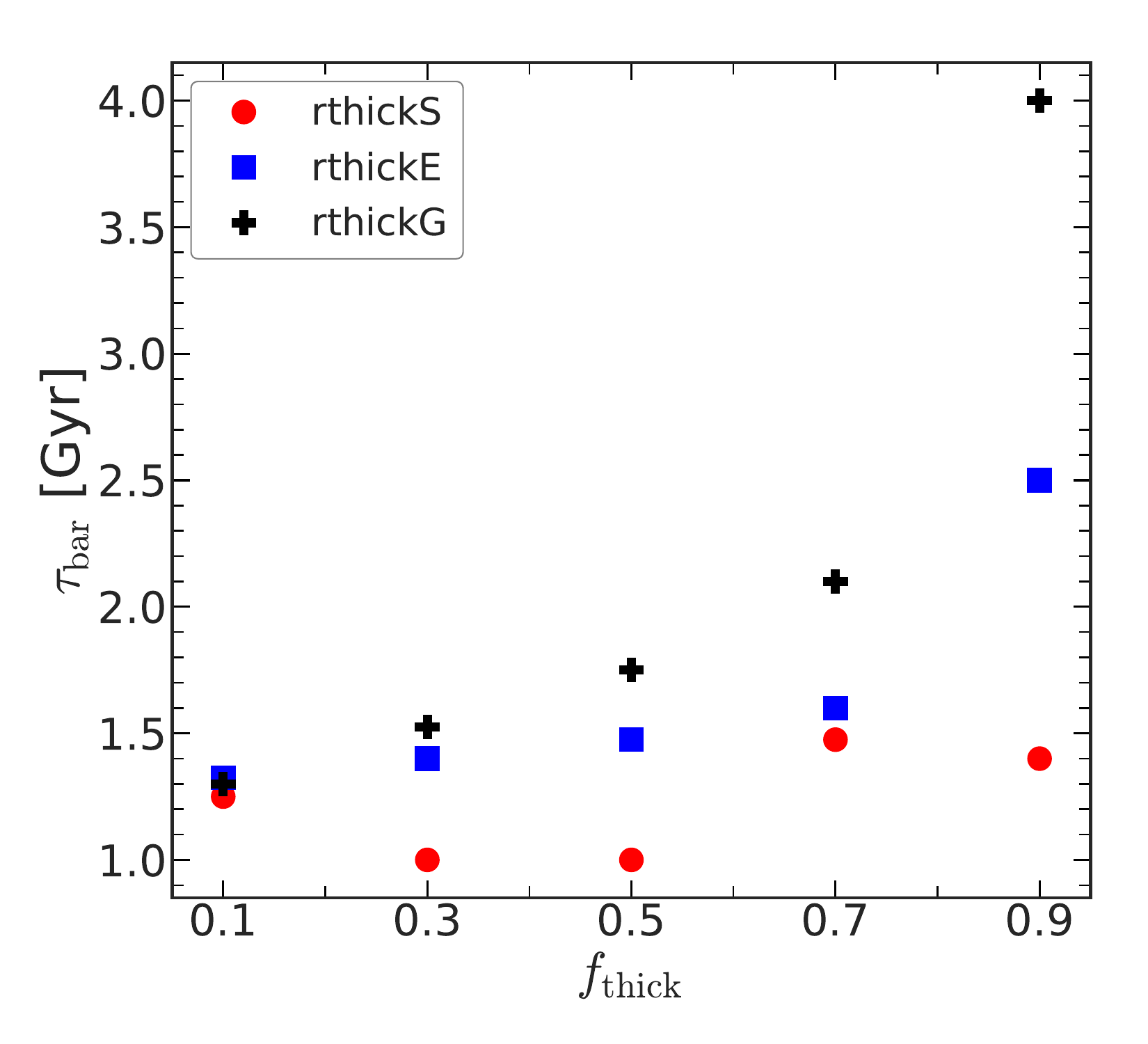}}
\caption{Variation of the bar formation epoch, $\tau_{\rm bar}$ as a function of thick-disc mass fraction ($f_{\rm thick}$), for all thin+thick models considered here. For a given geometric configuration, the bar formation epoch increases with increasing $f_{\rm thick}$ values, and this holds true for all three geometric configurations considered here.}
\label{fig:bar_formation_delay}
\end{figure}
%##################
% End figure
%##################

In \citet{Ghoshetal2022}, we showed that the bars in rthickS models tend to form earlier when compared to other two models (rthickE and rthickG). However, the bar formation epoch has not been quantified in \citet{Ghoshetal2022}. Here, we study this and also investigate how it varies with the thick-disc mass fraction. The bar formation epoch, $\tau_{\rm bar}$ is defined when the amplitude of the $m=2$ Fourier moment becomes greater than 0.2 and the corresponding phase angle, $\phi_2$ remains constant (within $3-5 \degrees$) within the extent of the bar. In Fig.~\ref{fig:bar_formation_delay}, we show the corresponding variation of the bar formation epoch, $\tau_{\rm bar}$ as a function of the thick-disc mass fraction, for all three geometric configurations considered here. As seen clearly, for a fixed $f_{\rm thick}$ value, bars form at an earlier epoch in rthickS models as compared to other two configurations. In addition, for a given geometric configuration, the bar formation epoch gets progressively increased with increasing $f_{\rm thick}$ values.

\section{Vertical-to-radial velocity dispersion profile}
\label{appen:Veldisp_example}
%&&&&&&&&&&&&&&&&&&&&&&&&&&&&&&&&&&&&&&&&&&&&&

%##################
% Begin figure
%##################
\begin{figure*}
\centering
\resizebox{0.95\linewidth}{!}{\includegraphics{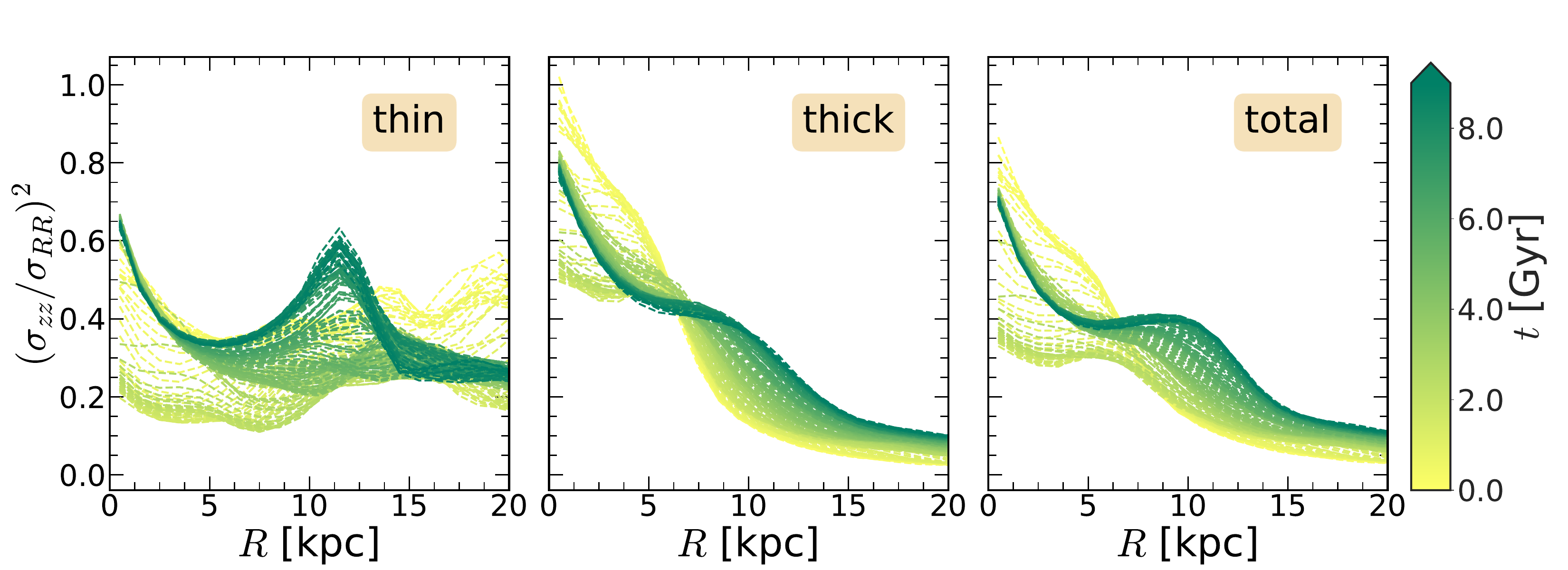}}
\caption{Radial profiles of the vertical-to-radial velocity dispersion ($\sigma_{zz}/\sigma_{RR}$), for the thin disc (left panels), thick disc stars (middle panels), and total (thin+thick) disc stars (right panels) as a function of time (shown in colour bar) for the model rthickE0.5.}
\label{fig:radialsigmazoversigmaR_profiles_rthickE05}
\end{figure*}
%##################
% End figure
%##################

%
%##################
% Begin figure
%##################
\begin{figure*}
\centering
\resizebox{\linewidth}{!}{\includegraphics{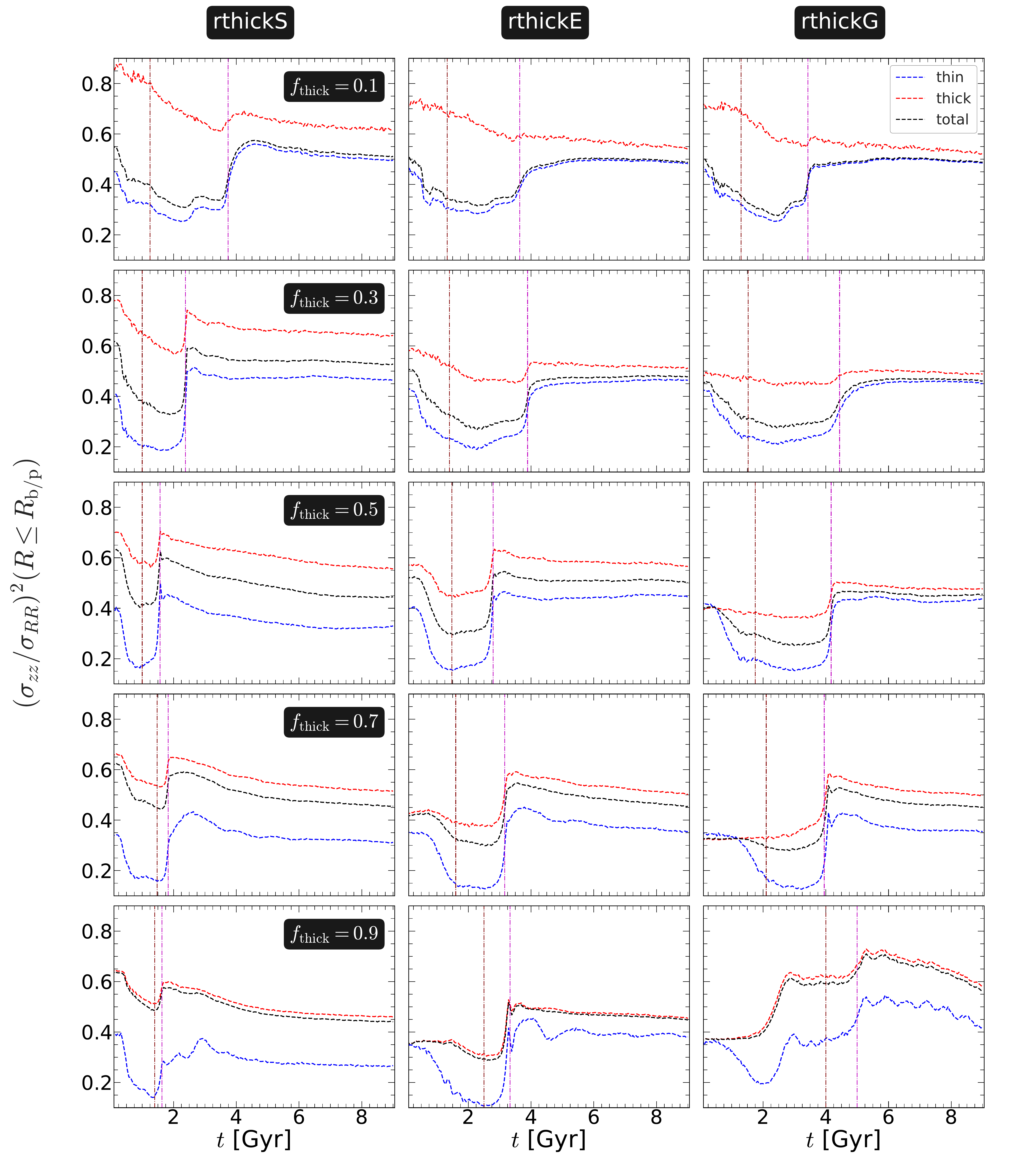}}
\caption{Temporal evolution of the vertical-to-radial velocity dispersion ($\sigma_{zz}/\sigma_{RR}$), calculated within $R_{\rm b/p}$, $(\sigma_{zz}/\sigma_{RR})^2 (t; R \leq R_{\rm b/p})$, for the thin (in blue), thick (in red), and total (thin+thick) disc (in black) particles, for all thin+thick disc models. \textit{Left panels} show the evolution for the rthickS models whereas  \textit{middle panels} and \textit{right panels} show the evolution for the rthickE  and rthickG models, respectively. The thick disc fraction ($f_{\rm thick}$) varies from 0.1 to 1 (top to bottom), as indicated in the left-most panel of each row. The vertical maroon dotted line denotes the onset of bar formation ($\tau_{\rm bar}$) while the vertical magenta dotted line denotes the onset of buckling instability ($\tau_{\rm buck}$). For details, see the text.}
\label{fig:rsigmazoversigmaR_temporal_allmodels}
\end{figure*}
%
%##################
% End figure
%##################

Fig.~\ref{fig:radialsigmazoversigmaR_profiles_rthickE05} shows the radial profiles of the vertical-to-radial velocity dispersion, $\sigma_{zz}/\sigma_{RR}$, as a function of time, for the thin+thick model rthickE0.5. Within the central $15 \kpc$ region, encompassing the b/p structure, the $\sigma_{zz}/\sigma_{RR}$ profiles show a huge variation as a function of time, especially seen for the thin disc stars. This time variation in radial profiles of $\sigma_{zz}/\sigma_{RR}$ are seen for other thin+thick models as well which subsequently formed a prominent b/p structure. In addition, in Fig.~\ref{fig:rsigmazoversigmaR_temporal_allmodels}, we show temporal evolution of the vertical-to-radial velocity dispersion ($\sigma_{zz} / \sigma_{RR}$), calculated separately for thin, thick, and thin+thick particles, for all thin+thick models considered here. The temporal evolution of $\sigma_{zz} / \sigma_{RR}$ displays a characteristic `U'-shape (of different amplitudes) during the course of the evolution, similar to what is seen for the model rthickE0.5. Furthermore, for a fixed value of $f_{\rm thick}$, the epoch corresponding to the minimum value of the quantity $\sigma_{zz} / \sigma_{RR}$ gets progressively delayed as the thick disc scale length increases (from rthickS models to rthickG models). This is in agreement with the finding that in  rthickG models, the b/p structure forms at a later time when compared to the other two configurations considered here (for details, see sect.~\ref{sec:bp_properties}.)

\end{appendix}

\end{document}